\documentclass[aps,reprint,nofootinbib]{revtex4-1}
\usepackage[latin1]{inputenc}
\usepackage{amsmath}
\usepackage{amsfonts}
\usepackage{amssymb}
\usepackage{graphicx}
\usepackage{blindtext}
\usepackage{color}
\usepackage{natbib}

\def\be{\begin{equation}}
\def\ee{\end{equation}}
\def\Ap{\theta_{V}}
\def\Al{\theta_{\ell}}
\def\AA{\phi}
\def\Sin{\text{sin}}
\def\Cos{\text{cos}}
\def\bea{\begin{eqnarray}}
\def\eea{\end{eqnarray}}
\newcommand{\nn}{\nonumber}

\begin{document}
	
			\title{Predictions of Angular Observables for $\bar{B}_s\to K^{\ast}\ell\ell$ and $\bar{B}\to \rho\ell\ell$  in Standard Model }
	\author{Bharti Kindra}
	\email{bharti@prl.res.in}
	\affiliation{Physical Research Laboratory, Navrangpura, Ahmedabad 380 009, Gujarat, India}
	\affiliation{Indian Institute of Technology Gandhinagar, Gandhinagar 382 424, Gujarat, India}
	\author{Namit Mahajan}
	\email{nmahajan@prl.res.in}
	\affiliation{Physical Research Laboratory, Navrangpura, Ahmedabad 380 009, Gujarat, India}
		\begin{abstract}
		Exclusive semileptonic decays based on $b\to s$ transitions have attracted a lot of attention as some angular observables deviate significantly from the standard model (SM) predictions. B meson decays induced by other flavor changing neutral current (FCNC),	$b\to d$, can also offer a probe to new physics with an additional sensitivity to the weak phase in the Cabibbo-Kobayashi-Masakawa (CKM) matrix. We provide predictions for angular observables for $b\to d$ semileptonic transitions, namely $\bar{B}_s\to K^{\ast}\ell^+\ell^-$, $\bar{B}^0\to\rho^0 \ell^+\ell^-$, and their CP-conjugated modes including various non-factorizable corrections. For $\bar{B}^0\to\rho^0 \ell^+\ell^-$ mode, $B^0-\bar{B}^0$ mixing effects have been included and predictions are made for Belle and LHCb separately. Study of these decay modes will be useful in its own right and to understand the pattern of deviations in $b\to s$ transitions.    
	\end{abstract}
	
	\maketitle

	\section{  Introduction}
	In recent years, a lot of attention has been given to semileptonic decays of bottom hadrons as a result of 
	increasing experimental evidence of new physics. Many decays have been observed involving the FCNC
	$b\to s\ell^+\ell^-$ and charged current $b\to c\ell \nu$. Most reliable measurements include
	$R_{K^{(*)}}$ \cite{LHCbrkstar,lhcrk} and $R_{D^{(*)}}$
	\cite{Babarrdstar1, Bellerd1, Bellerdstar1, LHCbrdstar1,Bellerdstar2} which hint towards 
	lepton flavor universality (LFU) violation. These measurements are important for precision tests of the standard model as well as for searches of new physics.\newline
	  	
   Albeit there exist rich data for $b\to s \ell^+\ell^-$ induced processes, the $b\to d$ counterpart of the weak
   decay, i.e. $b\to d\ell^+\ell^-$, has not caught much attention perhaps because of low branching ratio. 
   At the quark level, the lowest order contribution arises at one loop level through diagrams similar to $b\to s\ell \ell$ 
   which include box diagrams and electroweak penguin diagrams. The weak phases incorporate CKM matrix elements 
   $\xi_{q}^i=V_{qi}^* V_{qb}$, where $q\in \{u,c,t\}$ and $i\in\{s,d\}$. For $b\to s \ell\ell$ transition, $\xi^s_{c,t}\sim \lambda^2$ and $\xi^s_{u}\sim \lambda^4$ where $\lambda=0.22$. Since $u\bar{u}$ contribution introduces CKM phase which is negligible for $b\to s \ell \ell$, CP violating quantities are very small in SM. On the other hand, since $\xi^d_u\sim \xi^d_c\sim \xi^d_t\sim \lambda^4$ for $b\to d\ell \ell$, the B decays mediated through this transition allow for large CP violating quantities. Also, leading order contribution in this case is smaller than the leading contribution in $b\to s \ell \ell$ which makes this channel more sensitive to new physics. Hence, it is desirable to study processes like $B\to \{\pi,\rho\}\ell^+\ell^-$ and $B_s\to \{\bar{K},\bar{K}^{*}\}\ell^+\ell^-$ experimentally as well as theoretically. The first transition of this variety to be measured is $B\to \pi \ell^+\ell^-$ by LHCb with $5.2\sigma$ significance \cite{Btopiexp} which is in good agreement the expected value in SM \cite{Wang:2007sp,Hou:2014dza}. Other than this $B^0(B_s^0)\to \pi^+\pi^-\mu^+\mu^-$ has also been observed by LHCb, where the muons in final state do not originate from a resonance \cite{Aaij:2014lba}. In this paper, we focus on two decay modes: $B\to \rho\mu^+\mu^-$ and $B_s\to \bar{K}^{\ast}\mu^+\mu^-$. Predicted value of branching ratio for $B^0\to \rho^0 \ell^+ \ell^-$ is of the same order as $B\to \pi \ell^+\ell^-$, thus making it possible to be measured with upgraded experimental facilities. Since experiments already have good measurements of $B\to K^{\ast}\ell^+\ell^-$ mode, it is likely that $B_s\to \bar{K}^{\ast}\ell^+\ell^-$ mode may get early attention. The branching ratio of this mode is expected to be a factor of two more that $B^0\to \rho^0\ell^+\ell^-$ owing to the factor of $1/\sqrt{2}$ in the definition of $\rho^0\sim(u\bar{u}-d\bar{d})/\sqrt{2}$ as compared to $K^{\ast}\sim\bar{s}d$. Further, neglecting SU(3) breaking effects, one expects the branching fractions to be $\mathcal{O}(\lambda^2)$ smaller than those in $b\to s\ell^+\ell^-$ decays. Predictions for certain observables including branching ratio, direct CP asymmetry, and forward-backward asymmetry have been given for $B\to \rho\ell^+\ell^-$  \cite{Kruger:1999xa,Aliev:1998sk,Choudhury:2002fk,Beneke:2004dp,Faustov:2014zva}. For $B_s\to \bar{K}^{\ast}\mu^+\mu^-$, only branching ratio has been studied based on relativistic quark model \cite{Faustov:2013pca} and light cone sum rules (LCSR) based on heavy quark effective theory (HQET) approach \cite{Wu:2006rd}. However, a complete study of angular distribution is lacking. We aim to fill the gap in this paper.\newline
   
    Phenomenological analysis of the decays induced by this channel will provide complementary information about the nature of New Physics (NP).  The most prevailing problem in the theoretical description is due to the long distance effects of $c\bar{c}$ and $u\bar{u}$ resonant states. In the $q^2$ region close to these resonances, only model dependent predictions are available which result in large uncertainty. To avoid these uncertainties, we restrict our study to a region which  is well below $J/\psi$ resonance region $\sim 6 ~\text{GeV}^2$.\newline
              
     The paper is organised as follows. In section II we present the effective Hamiltonian for a general semileptonic B decay mediated by $b\to d$ transition at the quark level, and pseudoscalar to vector transition at hadronic level. In  section III, we discuss the form factors and inputs used to obtain numerical results. In section IV, we list all the observables considered in this paper. Results of these observables are given in section V followed by a summary in section VI. 
		
   \section{  Decay Amplitude}\label{effective}
	The theoretical description of semileptonic B decays is based on the effective Hamiltonian approach in which heavy degrees of freedom, i.e. top quark and gauge bosons, are integrated out. This approach allows the separation of short and long distance effects which are encoded in Wilson coefficients $\mathcal{C}_i$ and effective operators $\mathcal{O}_i$ respectively.  The effective Hamiltonian for $b\to d \ell^+\ell^-$ transitions, within SM, is expressed as \cite{balloperators}:
	 \begin{equation}\label{Hamiltonian}
	\mathcal{H}_{eff}=-\frac{4G_F}{\sqrt{2}}(\xi_t\mathcal{H}_{eff}^{(t)}+\xi_u\mathcal{H}_{eff}^{(u)})+ h.c.,
	 \end{equation}
	 where, $G_F$ is the Fermi constant, $\xi_q=V_{qd}^*V_{qb}^{\quad}$ are the CKM factors and,
	 \begin{subequations}\label{effective}
\begin{eqnarray}	 \mathcal{H}_{\text{eff}}^{{t}}&=&C_1 \mathcal{O}_1^c+C_2 \mathcal{O}_2^c+\sum_{i=3}^{10}C_i \mathcal{O}_i,\\
	 \mathcal{H}_{\text{eff}}^{{u}}&=&C_1 \left(\mathcal{O}_1^c-\mathcal{O}_1^u\right)+C_2\left(\mathcal{O}_2^c-\mathcal{O}_2^u\right).
	 \end{eqnarray}
 \end{subequations}  
where, the dominant contribution is due to the following operators:
	\begin{align}
	\mathcal{O}_1^u&=(\bar{d}_L\gamma_{\mu}T^au_L)(\bar{u}_L\gamma^{\mu}T^ab_L) \nn \\
	\mathcal{O}_2^u&=(\bar{d}_L\gamma_{\mu}u_L)(\bar{u}_L\gamma^{\mu}T^ab_L) \nn\\
	\mathcal{O}_1^c&=(\bar{d}_L\gamma_{\mu}T^ac_L)(\bar{c}_L\gamma^{\mu}T^ab_L)\nn\\
	\mathcal{O}_2^c&=(\bar{d}_L\gamma_{\mu}c_L)(\bar{c}_L\gamma^{\mu}b_L)\nn\\
	\mathcal{O}_7&= \frac{e}{g^2}m_b(\bar{d}_L\sigma^{\mu\nu}b_R)F_{\mu\nu}\nn\\
	\mathcal{O}_8&=\frac{1}{g} m_b (\bar{d}_L\sigma^{\mu\nu}T^ab_R)G^a_{\mu\nu}\nn\\
	\mathcal{O}_9&=\frac{e^2}{g^2}(\bar{d}_L\gamma_{\mu}b_L)(\bar{\ell}\gamma^{\mu}\ell)\nn\\
	\mathcal{O}_{10}&=\frac{e^2}{g^2}(\bar{d}_L\gamma_{\mu}b_L)(\bar{\ell}\gamma^{\mu}\gamma_5\ell).
	\end{align}
	Here, $T^a$ represents generators of $SU(3)$ group. In new physics scenario, operators other than given in Eq.(\ref{effective}) can also contribute significantly. The full operator basis for the effective Hamiltonian can be found in \cite{balloperators}. Unitarity of CKM matrix has been utilized to obtain Eq.(\ref{effective}). $\mathcal{H}_{\text{eff}}^t$ term contains the contribution of $t\bar{t}$ and $c\bar{c}$ quark-antiquark pair in the loop, while $\mathcal{H}_{\text{eff}}^u$ represents contribution of $c\bar{c}$ and $u\bar{u}$ pair in the loop. $\mathcal{C}_i$'s are the Wilson coefficients and are calculated at scale $\mu=m_W$ and expressed as a perturbative expansion in the strong coupling constant $\alpha_s(\mu_W)$:
\begin{align}
C_i(\mu_W)=& C_i^{(0)}(\mu_W)+\frac{\alpha_s(\mu_W)}{4\pi}C_i^{(1)}(\mu_W)\nonumber\\
& \left(\frac{\alpha_s(\mu_W)}{4\pi}\right)^2C_i^{(2)}(\mu_W)+...
\end{align} 
The Wilson coefficients have been worked out in \cite{Asatrian:2001de,Beneke:2001at,Asatrian:2003vq,Gorbahn:2004my,Bobeth:1999mk,Asatryan:2001zw} upto next-to-next-to leading order (NNLO). They are then evolved down to scale $\mu=m_b$ using renormalization group equations which is again expressed as a series with $\alpha_s/(4\pi)$ as expansion parameter. 
This step requires a calculation of anomalous dimension matrix $\gamma(\alpha_s)$ upto three-loop level to compute $C_i(\mu_b)^{(2)}$ which has been computed in \cite{Gambino:2003zm}. To the NNLO approximation, Wilson coefficients are given as \cite{Asatrian:2003vq},
\begin{subequations}\label{EqWC}
		\begin{align}
C_9^{\text{eff}}&=(1+\frac{\alpha_s(\mu)}{\pi}\omega_9(\hat{s}))\Big{(}A_9-\frac{\xi_c}{\xi_t}T_{9a}~h(\hat{m_c}^2,\hat{s}) \nonumber\\ &-\frac{\xi_u}{\xi_t}T_{9a}~h(0,\hat{s})+T_{9b}~h(\hat{m_c}^2,\hat{s})+U_9~h(1,\hat{s})\nonumber\\&+W_9~h(0,\hat{s})\Big{)} + \frac{\alpha_s(\mu)}{4\pi} \Big{(}\frac{\xi_u}{\xi_t}(C_1^{(0)}F_{1,u}^{(9)}+C_2^{(0)}F_{2,u}^{(9)})\nonumber\\&+\frac{\xi_c}{\xi_t}(C_1^{(0)}F_{1,c}^{(9)}+C_2^{(0)}F_{2,c}^{(9)})-A_8^{(0)}F_8^{(9)}\Big{)},\\
C_7^{\text{eff}}&=(1+\frac{\alpha_s(\mu)}{\pi}\omega_7(s))~A_7\nonumber \\&+\frac{\alpha_s(\mu)}{4\pi} \Big{(}\frac{\xi_u}{\xi_t}(C_1^{(0)}F_{1,u}^{(7)}+C_2^{(0)}F_{2,u}^{(7)})\nonumber\\&+\frac{\xi_c}{\xi_t}(C_1^{(0)}F_{1,c}^{(7)}+C_2^{(0)}F_{2,c}^{(7)})-A_8^{(0)}F_8^{(7)}\Big{)}\\
C_{10}^{\text{eff}}&=(1+\frac{\alpha_s(\mu)}{\pi}\omega_9(\hat{s}))A_{10}
\end{align}
\end{subequations} 
where $\hat{s}=q^2/m_b^2$ is the momentum squared of the lepton pair normalized to squared mass of $b$ quark. One major difference, as compared to $b\to s\ell^+\ell^-$ transition, is the presence of a large $\xi_u/\xi_t$ term in the Wilson coefficients. In $b\to s$ transition, this term is negligible in comparison to other terms and can be conveniently neglected. This implies that the Wilson coefficients, say $C_9^{\text{eff}}$, receive a small imaginary contribution. But for $b\to d$ case, the imaginary part is quite significant. \newline

Using the effective Hamiltonian in Eq. (\ref{Hamiltonian}), the matrix element for the hadronic decay $P\to V \ell^+\ell^-$ is written as product of short-distance contributions through Wilson coefficients and long-distance contribution which is further expressed in terms of form factors,
\begin{align}\label{eqnHamil}
	\mathcal{M}=&\frac{G_F \alpha}{\sqrt{2}\pi}V_{tb}V_{td}^* \Big{\{}\Big[\left<V|\bar{d}\gamma^{\mu}(C_9^{\text{eff}}P_L)b|P\right> \nonumber\\&-\frac{2m_b}{q^2}\left<V|\bar{d}~i~ \sigma^{\mu\nu}q_{\nu}(C_{7}^{\text{eff}}P_R)b|P\right>\Big](\bar{\ell}\gamma_{\mu}\ell)\nonumber\\
	&+\left<V|\bar{d}\gamma^{\mu}(C_{10}^{\text{eff}}P_L)b|P\right>(\bar{\ell}\gamma_{\mu}\gamma_5\ell)-16 \pi^2\frac{\bar{\ell}\gamma^{\mu}\ell}{q^2} \mathcal{H}_{\mu}^{\text{non-fac}}\Big{\}} .
\end{align}	

\begin{table*}
	\begin{center}
		\begin{tabular}{|ll|}
			\hline 
			 \multicolumn{2}{|c|}{Meson decay constants}\\ \hline
			$f_B=200\pm30$MeV , $f_{\rho,\perp}=150\pm 25$MeV	,$f_{\rho,\parallel}=209\pm 1$MeV & \cite{Beneke:2004dp}\\
			$f_{B_s}=230\pm 30$MeV, $f_{K^{\ast},\perp}=175\pm25$ MeV, $f_{K^{\ast},\parallel}=218\pm4$ MeV& \cite{Beneke:2003zv}\\ \hline 
			  \multicolumn{2}{|c|}{Couplings of meson distribution amplitudes}\\ \hline
			$a^{\parallel}_{1,K^{\ast}}=0.17\pm 0.04$, $a^{\parallel}_{2,K^{\ast}}=0.05\pm 0.05$,$a^{\parallel}_{1,\bar{K}^{\ast}}=-0.17\pm 0.04$, $a^{\parallel}_{2,\bar{K}^{\ast}}=0.05\pm 0.05$,& \cite{Ball:1998fj} \\
			$a^{\perp}_{1,K^{\ast}}=0.18\pm 0.05$, $a^{\perp}_{2,K^{\ast}}=0.03\pm 0.03$,$a^{\perp}_{1,\bar{K}^{\ast}}=-0.18\pm 0.05$, $a^{\perp}_{2,\bar{K}^{\ast}}=0.05\pm 0.05$,& \\ \hline
			\multicolumn{2}{|c|}{Wilson coefficients}\\ \hline
			$C_1=-0.255$, $C_2=1.014$, $C_3=-0.0048$, $C_4=-0.078$, $C_5=0.0003$, $C_6=0.0009$& \cite{Blake:2016olu}\\
			$C_7^{\text{eff}}=-0.2902$, $C_8^{\text{eff}}=-0.1599$, $C_9=4.033$, $C_{10}=-4.187$&\\ \hline
		\end{tabular}
		\caption{\label{input}Summary of input parameters used to calculate non-factorizable corrections.  }
	\end{center}
\end{table*}
where, $\mathcal{H}_{\mu}^{\text{non-fac}}$ represents the non-factorizable contribution of non-local hadronic matrix element. This results from four quark and chromomagnetic operators combined with virtual photon emission which then decays to lepton pair. These corrections are given in terms of hard-scattering kernels ($\mathcal{T}_a^q s$), where $a\in \{\perp,\parallel\}$ and $q\in \{u,c\}$, which are convoluted with B(B$_s$)-meson and $\rho(\bar{K^{\ast}})$ distribution amplitudes. The non-factorizable corrections included here are discussed below.\newline

\textit{Spectator Scattering}: The intermediate quark loop (up or charm) or chromomagnetic operator ($O_{8g}$) can emit a hard gluon which can be absorbed by the spectator quark. The $\mathcal{T}_a^{q,\text{spec}}$ for spectator scattering are obtained from Eqs.(20,22) of \cite{Beneke:2001at} and Eqs.(49,50) of \cite{Beneke:2004dp}.\newline

 \textit{Weak Annihilation}: The b quark in the B meson can annihilate with the spectator quark to give the meson in the final state. This contributes to the leading order term in $\alpha_s$ as the QCD correction to weak annihilation is highly suppressed. The non-zero contribution of weak annihialtion ($\mathcal{T}_{\parallel}^{q,\text{WA}}$) is given by Eqs. (46-48) of \cite{Beneke:2004dp}.\newline 
 
  \textit{Soft-gluon correction}: Quark in the intermediate loop can emit a soft gluon which contributes to non-factorizable correction. The contribution is proportional to $1/(4m_c^2-q^2)$, and rises near the vector resonances. Hence, it can be calculated in the region $[2-6]$GeV$^2$. In the region beyond that, hadronic dispersion relations are employed which systematically includes the contribution of charm resonance \cite{charmloop}. For $B\to K^{\ast}\ell\ell$ the contribution of these two effects, soft-gluon emission and charmonium resonances, can be described by the parameterization defined in Eqs. (8-12) of \cite{Descotes-Genon:2015uva} which is valid in the $q^2$ region $[1-9]$GeV$^2$. Such corrections have not been explicitly computed for $B\to \rho\ell\ell$ or $B_s\to \bar{K}^{\ast}\ell\ell$. However, flavour SU(3) symmetry would imply that these corrections can be assumed to be roughly same as $B\to K^{\ast}\ell\ell$, which are given by,
  \begin{align}
\Delta C_{9,c}^{\perp,\text{soft}}(q^2)& =\frac{a^{\perp}+b^{\perp} q^2(c^{\perp}-q^2)}{q^2(c^{\perp}-q^2)}\\
\Delta C_{9,c}^{\parallel,\text{soft}}(q^2)& =\frac{a^{\parallel}+b^{\parallel} q^2(c^{\parallel}-q^2)}{q^2(c^\parallel-q^2)}\\
\Delta C_{9,c}^{0,\text{soft}}(q^2)& =\frac{a^0+b^0 (q^2+1)(c^0-q^2)}{(q^2+1)(c^0-q^2)}
  \end{align}
where, the mean values of parameters are given in Table \ref{TabCharm}. 
\begin{table}\label{TabCharm}
	\begin{tabular}{|l|c|c|c|}
		\hline
	i~~~&	$a^i$~~& $b^i~~$& $c^i~~$\\ \hline
	$\perp$~&	9.25 &-0.5& 9.35\\
	$\parallel$& 9.25 &-0.5& 9.35\\
	$0$& 33 & -0.9 & 10.35 \\ \hline
	\end{tabular} \caption{Values of parameters defined for $\Delta C_9^{i,\text{soft}}$}
\end{table}
However, the expressions for soft-gluon emission from the up loop are still absent and need to be computed properly. Though the corresponding expressions exist for $B\to \pi \ell\ell$ mode but they can not be naively used for the present purpose. For current study, we are assuming an uncertainty of $\sim 10\%$ in $C_9$ to account for this missing piece:
\begin{equation}
\delta C_{9,u}^{\text{soft}}= a e^{i \theta}
\end{equation}
where, $|a|\in\{0,0.5\}$ and $\theta\in\{0,\pi\}$. The evaluation, particularly the sign, of this correction requires a complete LCSR calculation which is beyond the present work. The impact of these contributions doesn't turn out to be very significant except for one or two observables. However, to be complete and to indicate possible effect of these corrections, we include them in our numerical study.\newline

 These corrections are added systematically in transversity amplitudes which are given in Appendix \ref{appendixAmplitudes}.
 
\section{ Form Factors}\label{formfactors}
 The matrix elements corresponding to operators $\mathcal{O}_{7,9,10}$ are expressed in terms of seven form factors which are functions of $q^2$:
\begin{eqnarray}\label{eq:B2K2ff}
&&c_V\langle V(k)|\bar d\gamma^{\mu}b|B(p)\rangle
=\frac{2V(q^2)}{m_B+m_{V}}\epsilon^{\mu\nu\rho\sigma} \epsilon^*_{\nu}  p_{\rho}k_{\sigma}, \nonumber\\&&
c_V\langle V(k)|\bar d\gamma^{\mu}\gamma_5 b|P(p)\rangle=2im_{V} A_0(q^2)\frac{\epsilon^*\cdot  q }{ q^2}q^{\mu}\nn\\&& ~~~~~~~~~+ i(m_B+m_{V})A_1(q^2)\left[ \epsilon^*_{\mu }
-\frac{\epsilon^* \cdot  q }{q^2}q^{\mu} \right] \nonumber\\
&&~~~~~~~~-iA_2(q^2)\frac{\epsilon^* \cdot  q }{  m_B+m_{V} }
\Big[ P^{\mu}-\frac{m_B^2-m_{V}^2}{q^2}q^{\mu} \Big],\nn\\&&
c_V\langle   V(k)|\bar dq_{\nu}\sigma^{\mu\nu}b|P(p)\rangle
=2iT_1(q^2)\epsilon^{\mu\nu\rho\sigma} \epsilon^*_{\nu} p_{\rho}k_{\sigma}, \nonumber\\&&
c_V\langle  V(k)|\bar dq_{\nu}\sigma^{\mu\nu}\gamma_5b|P(p)\rangle
=T_2(q^2)\Big[(m_B^2-m_{V}^2) \epsilon^*_{\mu }\nn\\&&~~~~
- {\epsilon^* \cdot  q }  P^{\mu} \Big]+T_3(q^2) {\epsilon^* \cdot  q }\left[
q^{\mu}-\frac{q^2 (p+k)^{\mu}}{m_B^2-m_{V}^2}\right],
\end{eqnarray}
where $q_{\mu}=(p-k)_{\mu}$, $P_{\mu}=(p+k)_{\mu}$, and $c_V=1/\sqrt{2}$ in the case of $\bar{B}^0\to \rho^0\ell\ell$; 1 for $\bar{B}_s\to K^{\ast}\ell\ell$ and $\bar{B}^{\pm}\to \rho^{\pm}\ell\ell$. Form factors can be calculated using the
method of QCD Sum Rules on Light-Cone (LCSRs) in the low-$q^2$ region. For semileptonic B decays, the method involves calculation of correlation function of the weak currents involving $b$ quark, evaluated between 
the vacuum and light meson in the final state. The correlation function is factorized into non-perturbative and process-independent hadron distribution amplitudes (DAs), $\phi$, convoluted with  process-dependent amplitudes $T$.
	\begin{equation}
\text{correlation function}\sim \sum_n T^{(n)}\otimes \phi^{(n)}
\end{equation}
	where, $n$ represents twist. The contributions with increasing twist decreases by increasing powers of virtualities of the currents involved ($\sim m_b^2$ in the low $q^2$ range). We follow \cite{updated} for form factors of $B\to \rho$ and $B_s\to K^{\ast}$ hadronic decays, which provides an improved determination of $B\to V $ form factors compared to those in \cite{formfactors}. In \cite{updated}, updated values of hadronic parameters are used and contributions upto twist-5 in DAs have been systematically included. Further, making use of equations of motion, it is shown that the uncertainties in the ratios of form factors are reduced and so does the dependence on mass scheme. Another advantage is that the combined fits to sum rules and lattice calculations at low and high $q^2$ are given which provides form factors valid over the whole range. In this paper, we call the updated form factors as BSZ (Bharucha-Straub-Zwicky) form factors while those in \cite{formfactors} as BZ (Ball-Zwicky) form factors.\newline
	
	 The form factors are written as a series expansion in terms of the parameter \cite{updated},
	\begin{equation}
	z(t)=\frac{\sqrt{t_{+}-t}-\sqrt{t_{+}-t_0}}{\sqrt{t_{+}-t}+\sqrt{t_{+}-t_0}}
	\end{equation}
	where, $t_{\pm}=(M_B\pm M_V)^2$ and $t_0=t_{+}(1-\sqrt{1-t_{-}/t_{+}})$. Form factors are parameterized as:
	\begin{equation}
	F_{i}(q^2)=(1-s/m_{R,i}^2)^{-1}\Sigma_k \alpha_k^{i}\left[z(s)-z(0)\right]^k.
	\end{equation}
where $m_{R,i}$ is the resonance mass which is equal to $5.279$ Gev for $A_0(s)$, 5.325 GeV for $T_1(s)$ and $V(s)$, and 5.724 GeV for rest of the form factors. \newline
\begin{table}
		\begin{center}
			\begin{tabular}{|l|l|}
				\hline
				$M_B$=5.27 GeV &  $M_{\rho}$= 0.775$\pm$ 0.025 GeV\\
				$M_{B_s}$=5.366 GeV & $M_{K^{\ast}}=0.891$ GeV\\ 
				$m_b$=4.80$\pm$0.06 GeV & $m_c$=1.4 $\pm$ 0.2 GeV\\
				$\mu$=5 GeV & $\alpha_s$=0.215\\
				$G_F$=1.16 $\times$ 10$^{-5}$ GeV$^{-2}$ & $\alpha_{em}$=1/137\\
				$\lambda$=0.22506$\pm$0.00050 & A=0.811
				$\pm$ 0.026\\
				$\bar{\rho}$=0.124$^{0.019}_{-0.018}$ & $\bar{\eta}$=0.356$\pm$0.011
				\\ \hline
			\end{tabular}
		\caption{\label{input2}Input values used to generate values of observables }
		\end{center}
	\end{table}

Below, we provide detailed SM prediction employing BSZ form factors , computed using LCSRs, which we will refer as BSZ1 form factors in this paper. To compare the numerical impact of the improved 
form factors, we also provide a direct comparison with results obtained using BSZ form factors with lattice and LCSR results combined together (referred as BSZ2 form factors in this paper), and BZ form factors, in the case of $\bar{B}_s\to K^{\ast}\ell^+\ell^-$. While for $\bar{B}^0\to\rho^0\ell^+\ell^-$, we use BSZ form factors (LCSR) only, since combined fit with lattice results are not available for this mode. 
\section{ Observables}\label{observable}

   For a four body decay, $B\to V(\to M_1M_2)\ell^+\ell^-$, the decay distribution can be completely described in terms of four kinematic variables; 
    the lepton invariant mass squared ($q^2$) and three angles $\Ap$, $\theta_l$, and $\phi$. 
    The angle $\Ap$ is the angle between direction of flight of $M_2$ with respect to $B$ meson in the rest
    frame of $V$, $\Al$ is the angle made by $\ell^-$ with respect to the $B$ meson in the dilepton rest
    frame and $\phi$ is the azimuthal angle between the two planes formed by dilepton and $M_1M_2$. 
    The full angular decay distribution of $B\to V(\to M_1M_2)\ell^+\ell^-$ is given by \cite{balloperators},

\begin{align}\label{EQdistri}
&\frac{d^4\Gamma}{dq^2~dcos\Ap~d\theta_{\ell}~d\phi}=\frac{9}{32 \pi}I(q^2,\Ap,\Al,\AA)\end{align}
where,
\begin{align}
&I(q^2,\Ap,\Al,\AA)=\Big{(}I_1^s~ \Sin^2\Ap+I_1^c~\Cos^2\Ap\nn\\&+(I_2^s~\Sin^2\Ap+I_2^c~\Cos^2\Ap)\Cos2\Al+I_3~\Sin^2\Ap \Sin^2\Al \Cos2\AA\nn\\&+I_4 ~\Sin 2 \Ap \Sin 2\Al \Cos\AA+I_5~\Sin 2\Ap\Sin \Al\Cos\AA\nonumber \\ & +(I_6^s~\Sin^2\Ap+I_6^c~\Cos^2\Ap)\Cos\Al+I_7~\Sin 2\Ap\Sin\Al\Sin\AA \nn\\&+I_8~\Sin 2\Ap\Sin 2\Al \Sin\AA+I_9~\Sin^2\Ap\Sin^2\Al\Sin 2\AA\Big{)}.
\end{align}
Here, V is an intermediate vector meson which decays to $M_1$ and $M_2$ whereas $\ell^+\ell^-$ can be any lepton pair. The corresponding angular decay distribution ($d^4\bar{\Gamma}/(dq^2dcos\theta_{\pi}d\theta_{\ell}d\phi)$) for the 
CP-conjugated process,$ \bar{B}\to\bar{V}(\to\bar{M}_1\bar{M}_2)\ell^+\ell^-$, is obtained from Eq. (\ref{EQdistri}) 
with the replacement, $I_i\to \tilde{I}_i\equiv \zeta_i \bar{I}_i$, where, $\zeta_i=1$ for $i\in\{1,2,3,4,7\}$ and -1 
for $i\in \{5,6,8,9\}$. $\bar{I}_i$ is equal to $I_i$ with the weak phase, i.e. CKM phase in this case, conjugated. 
The functions $I_i$ can be written in terms of transversity amplitudes \cite{balloperators}. In the $b\to s$ transition, 
since the Wilson coefficients are effectively real, modulo a small imaginary part coming due to function $h(m^2,s)$ in
$C_9^{\text{eff}}$, $\bar{I}_i$ are essentially $I_i$ and observables sensitive to imaginary part of $I_i$ are rather
small within SM. This is not the case in $b\to d$ induced decays and we see this feature explicitly in the results below.
Various observables are constructed from Eq. (\ref{EQdistri}) by integrating over angles in various range. 
These observables are generally plagued with large uncertainties due to form factors. 
To avoid this, a lot of work has been done to construct observables which are theoretically clean 
in low-$q^2$ region \cite{implications,Kruger:2005ep,Egede:2008uy,Matias:2012xw,Egede:2010zc,Becirevic:2011bp,FFIVirto}. 
Such observables are free from this dependence at the leading order and are called \textit{form factor independent} 
(FFI) observables. Those which have a form factor dependence in the leading order are called \textit{form factor dependent}
(FFD) observables. We study both classes of observables in this paper, as discussed below. We shall see below, SU(3) breaking effects are clearly visible in some of the observables.\newline

\textbullet~ FFD observables are (which have been experimentally studied in the context of $B\to \bar{K}^{\ast}\ell\ell$\cite{Aubert:2006vb}):
\begin{subequations}\label{FFD}
	\begin{align}
	\frac{d\Gamma}{dq^2}=&\int_{-1}^{1}d\Cos\Al\int_{-1}^{1}d\Cos\Ap\int_{0}^{2\pi}\AA\frac{d^4\Gamma}{dq^2d\Cos\Ap d\Cos\Al d\AA}\nn\\&=\frac{1}{4}(3I_1^c+6I_1^s-I_2^c-2I_2^s)\\
	A_{FB}(q^2)=&\frac{1}{d\Gamma/dq^2}\Big[\int_{-1}^{0}-\int_{0}^{1}\Big]d\Cos\Al\frac{d^4\Gamma}{dq^2 d\Cos\Al}\nn\\ &=~\frac{-3I_6^s}{3I_1^c+6I_1^s-I_2^c-2I_2^s}\\
	F_L(q^2)=&~~\frac{3I_1^c-I_2^c}{3I_1^c+6I_1^s-I_2^c-2I_2^s} 
				\end{align}
\end{subequations}
where,$\frac{d\Gamma}{dq^2}$ is the dilepton spectrum distribution, $A_{FB}(q^2)$ is the forward-backward asymmetry and $F_L(q^2)$ is the fraction of longitudinal polarization of the intermediate vector meson. Similar observables are constructed for the CP-conjugate process using the decay distribution $d^4\bar{\Gamma}/(dq^2dcos\theta_{\pi}d\theta_{\ell}d\phi)$ discussed above. \newline

\textbullet ~ FFI observables or "clean observables" are independent of form factors in the leading order of $1/m_b$ and $\alpha_s$ thus exhibiting low hadronic uncertainties and enhanced sensitivity to new physics. Much attention has been given to the construction of such observables and some of them have been measured experimentally \cite{LHCbFFI,BelleFFI}. We consider following set of FFI observables here:
\begin{align}\label{FFI}
P_1&=\frac{I_3}{2 I_{2}^s},~ P_2=\beta_l\frac{I_6^s}{8I_2^s},~P_3=\frac{I_9}{4I_2^s},~
 P_4^{\prime}=\frac{I_4}{\sqrt{-I_2^cI_2^s}}\nonumber\\ P_5^{\prime}&=\frac{I_5}{2\sqrt{-I_2^cI_2^s}},~  
 P_6^{\prime}=-\frac{I_7}{2\sqrt{-I_2^cI_2^s}},~P_8^{\prime}=-\frac{I_8}{2\sqrt{-I_2^cI_2^s}}
\end{align}
\textbullet~ We also consider observables analogous to $R_{K^{\ast}}$ for which the form factor dependence cancels exactly for $B_s\to \bar{K}^{\ast}\ell\ell$, defined as:
\begin{equation}\label{LFUV}
R_{K^{\ast}}^{B_s}=\frac{\left[\mathcal{BR}(B_s\to\bar{K}^{\ast}\mu^+\mu^-)\right]_{q^2\in\{q_1^2,q_2^2\}}}{\left[\mathcal{BR}(B_s\to\bar{K}^{\ast}e^+e^-\right]_{q^2\in\{q_1^2,q_2^2\}}}
\end{equation}
where, numerator and denominator are integrated over $q^2$ in the range $ [q_1^2-q_2^2]$ GeV$^2$. Observables defined in Eqs. (\ref{FFD},\ref{FFI},\ref{LFUV}) are valid for $B_s\to\bar{K}^{\ast}\ell^+\ell^-$ decay mode. 
It has been pointed out in literature that the zeroes (value of $ q^2$ where observables is zero) are also clean observables \cite{Ali:1999mm,Kumar:2014bna}. Also, the relation between zeroes of different observables provide crucial tests of Standard Model. Thus, we also provide values of zeroes of different observables. Observables for the CP-conjugate decay, $\bar{B}_s\to K^{\ast}\ell^+\ell^-$ are also defined in the same way, with the substitution $I_i\to \bar{I}_i \equiv \zeta_i\tilde{I}_i$. Results for $B_s(\bar{B}_s)\to \bar{K}^*(K^*)\mu^+\mu-$ are given in the next section which can be compared with data collected at LHCb as well as Belle. \newline

However, for $B^0\to \rho^0 \ell^+\ell^-$, results corresponding to LHCb and Belle have to be computed separately. Since $\rho^0\to\pi\pi$, which is not a flavor specific state, the observables are affected by $B^0-\bar{B}^0$ oscillations and the expressions of angular functions ($I_i$s) defined in Eq. (\ref{EQdistri}), are modified. These modified, time-dependent functions have been computed in \cite{mixing} and given as, 
\begin{eqnarray}
J_i(t)+\tilde{J}_i(t)&=&e^{-\Gamma t}[(I_i+\tilde{I}_i)~\text{cosh}(y\Gamma t)\\ \nonumber & &-h_i~\text{sinh}(y\Gamma t)],\\
J_i(t)-\tilde{J}_i(t)&=&e^{-\Gamma t}[(I_i-\tilde{I}_i)~\text{cos}(x\Gamma t)\\ \nonumber & &-s_i~\text{sin}(x\Gamma t)],
\end{eqnarray}
where, $x=\Delta m/\Gamma$, $y=\Delta\Gamma/\Gamma$, $\tilde{J}_i\equiv \zeta_i J_i$, and the additional functions ($h_i$ and $s_i$) arise because of the mixing in $B^0$ meson system. This leads to two types of quantities, time-dependent observables and time-integrated observables. In this work, we consider observables which include time-integrated angular functions over a range $t\in [0,\infty)$ in the case of LHCb and $t\in ( -\infty,\infty)$(in addition to $\text{exp}(-\Gamma t) \to \text{exp}(-\Gamma |t|)$) at Belle \cite{babarbook,integration}. After time-integration, the modified angular functions are given by, 
\begin{eqnarray}\label{modified}
\left<J_i+\tilde{J}_i\right>_{\text{LHCb}}& =&\frac{1}{\Gamma}\left[\frac{I_i+\tilde{I}_i}{1-y^2} -\frac{y}{1-y^2}\times h_i\right],\\
\left<J_i-\tilde{J}_i\right>_{\text{LHCb}}& =&\frac{1}{\Gamma}\left[\frac{I_i-\tilde{I}_i}{1+x^2} -\frac{x}{1+x^2}\times s_i\right],\\
\left<J_i+\tilde{J}_i\right>_{\text{Belle}}& =&\frac{2}{\Gamma}\left[\frac{1}{1-y^2}\times (I_i+\tilde{I}_i)\right],\\
\left<J_i-\tilde{J}_i\right>_{\text{Belle}}& =&\frac{2}{\Gamma}\left[\frac{1}{1+x^2}\times (I_i-\tilde{I}_i)\right],
\end{eqnarray} 
where, $\left<~\right>$ represents time-integrated quantity. Other difference at LHCb and Belle arises due to the fact that flavor of the meson can be tagged using flavor-specific decays at Belle. Thus, flavor of the meson decaying to the final state is known at time $t=0$ and the appropriate angular function ($J_i$ or $\tilde{J_i}$) can be used. On the other hand, there is no method to determine the flavor of meson at $t=0$ at LHCb. As a result, the measured quantity at LHCb is $d\Gamma(B^0\to \rho^0\ell^+\ell^-)+d\bar{\Gamma}(\bar{B}^0\to \rho^0\ell^+\ell^-)$ allowing the observation of $\left<J_i+\tilde{J}_i\right>$ combination only, which is a CP-averaged quantity for $i\in\{1,2,3,4,7\}$ and CP-violating quantity for $i\in \{5,6,8,9\}$. Due to the difference in the method of measurement, we consider  different observables to be studied at LHCb and Belle.\newline
 
 For $B^0\to \rho^0\ell^+\ell^-$ at Belle, the definition of observables (say, O) in Eqs. (\ref{FFD},\ref{FFI},\ref{LFUV}) are modified as, $I_i \to\left<J_i\right>$ and are denoted by $\left<O\right>^{\text{Tagged}}$. Similarly, for the CP conjugate process, the observables $\left<\tilde{O}\right>^{\text{Tagged}}$ are obtained by modification $\bar{I}_i\to \tilde{J}_i$. For untagged events at LHCb, the required modification in the definition of observables is $I_i\to \left<J_i+\tilde{J}_i\right>$ and the observables are denoted as $\left<O\right>^{\text{Untagged}}$. To avoid any ambiguity, we have provided the definitions of full set of observables in Appendix \ref{appendixA}. It is clear from the form of observables that, Belle allows a study of angular distribution of both $B^0\to\rho^0\ell^+\ell^-$ and $\bar{B}^0\to\rho^0\ell^+\ell^-$ decays, while at LHCb, only a CP-averaged or CP-asymmetric study is possible. For the decays considered in this paper $y=0$. Thus, observables $\left<O\right>^{\text{Untagged}}$ are also measurable at Belle. Moreover, it can be noticed from Eq. (\ref{modified}), that the value of these observables ($\left<O\right>^{\text{Untagged}}$), if measured at Belle, are expected to be same at the two experiments, except $\left<BR\right>^{\text{LHCb}}$, which should be twice for Belle in comparison to LHCb.\newline
 
  Even though tagging power at LHCb is low, new algorithms have been suggested to improve the tagging power by $50\%$ \cite{Aaij:2016psi,Aaij:2016rdg}. Thus, for completion we also give predictions for observables which can be measured at LHCb using tagging of B mesons. The definition of these observables is again given by $\left<O\right>^{\text{Tagged}}$. Moreover, having measurements of angular distribution with and without tagging can be of phenomenological importance \cite{next}.

\section{Results}\label{results}
In this section, we present observables as a function of $q^2$ and their binned values over two $q^2$ ranges: [0.1-1] GeV$^2$ and [1-6] GeV$^2$ and consider the di-muon pair in the final state.  For $\bar{B}_s\to K^{\ast}\mu^+\mu^-$, we provide the results obtained using three sets of form factors. As discussed earlier, we mainly employ the BSZ(LCSR) form factors and compare the results obtained using BZ form factors and BSZ form factors (LCSR+Lattice results). For $B^0\to\rho^0\mu^+\mu^-$, we present results using the BSZ form factors only.\newline

We first discuss the observables without the inclusion of various non-factorizable corrections. After discussing the main results, we shall return to thr discussion of the impact of these corrections. In Table \ref{KstarwoCorrec}, we give values of observables for $B_s\to \bar{K}^{\ast}\mu^+\mu^-$ and $\bar{B}_s\to K^{\ast}\mu^+\mu^-$ corresponding to the form factor set BSZ1, BSZ2, and BZ. For $B(\bar{B})\to \rho \mu^+\mu^-$, the values of angular observables are given in Table \ref{rhowoCorrec} for the form factor set BZ. The errors in the binned values are due to errors in form factors as given in \cite{updated,formfactors}. \newline
\begin{table*}
	\begin{tabular}{|l||c|c|c|c|}
		\hline
		& \multicolumn{2}{c}{$\bar{B_s}\to K^{\ast}\mu^+\mu^-$} &  \multicolumn{2}{c|}{$B_s\to \bar{K}^{\ast}\mu^+\mu^-$}\\
		\hline
		Observable/ Bin& [0.1-1] GeV$^2$& [1-6] GeV$^2$&[0.1-1] GeV$^2$& [1-6] GeV$^2$ \\
		\hline
		$P_1$[BSZ1]& $0.006\pm0.132$ & $-0.081\pm0.129 $ & $0.005\pm0.131$ & $-0.071\pm 0.114$\\
		~~~[BSZ2] &$0.008\pm0.131$ & $-0.097\pm 0.128$ & $0.007\pm 0.129$&$-0.082\pm 0.113$\\
		~~~[BZ]& $0.006 $ & $-0.084$ & $0.005$ & $-0.069$ \\
		\hline
		
		$P_2$[BSZ1]& $0.124\pm0.013$& $0.011\pm0.078$& $0.111\pm0.011$& $0.094\pm0.076$\\
		~~~[BSZ2]& $0.117\pm0.013$ & $0.051\pm 0.081$& $0.104\pm 0.011$& $0.132\pm 0.075$\\
		~~~[BZ]& $0.127$ & $-0.012$ & $0.110$ & $0.095$ \\ \hline
		
		$P_3$[BSZ1]& $0\pm0.001$& $0.002\pm0.005$ & $0\pm0$& $0.001\pm0.002$\\
		~~~[BSZ2]& $0\pm 0.001$ & $0.002\pm0.005$& $0\pm 0$ & $0.001\pm0.002$ \\
		~~~[BZ]& $-0.0$ & $0.001$ & $0$ & $0$\\  \hline
		
		$P_4^{\prime}$[BSZ1]& $-0.488\pm0.053$& $0.619\pm0.151$ & $-0.489\pm0.051$& $0.526\pm0.159$\\
		~~~[BSZ2]& $-0.506\pm 0.052$ & $0.575 \pm 0.161 $ & $-0.508 \pm0.049$ & $0.473\pm 0.168$\\ 
		~~~[BZ]& $-0.489$ & $0.654$ & $-0.481$ & $0.537$\\ \hline
		
		$P_5^{\prime}$[BSZ1] & $0.633\pm0.057$& $-0.424\pm0.119$& $0.673\pm0.057$& $-0.324\pm0.125$\\
		~~~[BSZ2]& $0.635\pm0.055$& $-0.365\pm0.126$ & $0.659\pm0.053$ & $-0.264\pm 0.130$\\
		~~~[BZ]& $0.632$ & $-0.450$ & $0.679$ & $-0.328$ \\ \hline
		
		$P_6^{\prime}$[BSZ1] & $-0.098\pm0.006$& $-0.071\pm0.009$ & $0.004\pm0.001$& $-0.012\pm0.002$	\\ 
		~~~[BSZ2]& $-0.096\pm0.006$ & $-0.075\pm 0.010$& $0.004\pm0.001$ & $-0.013\pm 0.002$\\
		~~~[BZ]& $-0.0982$ & $-0.067$ & $0.004$& $0.024$  \\ \hline
		
		$P_8^{\prime}$[BSZ1] & $0.0234\pm 0.004$ & $0.023\pm 0.005$ & $0.003\pm0.001$ & $0.009\pm 0.002$ \\
		~~~[BSZ2]& $0.023\pm0.005 $ & $0.022\pm0.005$ & $0.003\pm0.001$ & $0.009\pm0.002$\\ 
		~~~[BZ]& $0.019$ & $0.017$ & $-0.005$ & $0.003$\\ \hline
		 
		$R_{K^{\ast}}^{B_s}$[BSZ1]& $0.939\pm 0.010$& $0.997\pm0.004$ & $0.935\pm 0.009$ & $0.997\pm0.004$\\
		~~~~~[BSZ2]&$0.944\pm 0.010$& $0.999\pm0.004$ & $0.939\pm 0.011$ & $0.998\pm0.040$\\
		~~~[BZ]& $0.929$ & $0.995$ & $0.932$ & $0.995$ \\	\hline
		
			$BR\times 10^{9}$[BSZ1]& $2.647\pm 0.331$ & $5.807\pm 1.418$& $3.159\pm0.378$ & $6.011\pm 1.452$\\
		~~~[BSZ2] & $3.019\pm0.366$ & $7.274\pm1.642$ & $3.526\pm0.409$ & $7.531\pm1.685$\\ 
		~~~[BZ]& $3.117$ & $7.107$ & $3.712$  & $7.329$\\ \hline
		
		$A_{FB}$[BSZ1] & $-0.078\pm0.009$& $-0.021\pm0.028$& $-0.076\pm 0.009$& $-0.033\pm 0.029$\\
		~~~[BSZ2] & $-0.065\pm0.008$ & $-0.012\pm0.021$ & $-0.065\pm0.001$ & $-0.035\pm0.021$\\
		~~~[BZ]& $-0.078$ & $0.004$ & $-0.071$& $-0.033$\\ \hline
		
		$F_L$[BSZ1]& $0.343\pm0.065$ & $0.824\pm0.050$ & $0.276\pm0.058$ & $0.800\pm0.053$\\
		~~~[BSZ2] & $0.414\pm 0.101$ & $0.876\pm0.356$ & $0.345\pm0.058$ & $0.848\pm0.038$ \\ 
		~~~[BZ]& $0.341$ & $0.841$ & $0.273$ & $0.815$\\
		\hline
	\end{tabular}
	\caption{\label{KstarwoCorrec}Binned values of observables for the process $\bar{B_s}\to K^{\ast}\mu^+\mu^-$ and $B_s\to \bar{K^{\ast}}\mu^+\mu^-$} are given for all the three sets of form factors. The uncertainties shown are due to errors in determination of form factors. 
\end{table*}

\begin{table*}
	\begin{tabular}{|l||c|c|c|c|c|c|}
		\hline
		& \multicolumn{2}{c}{$B\to\rho\mu^+\mu^-$}(Belle) &  \multicolumn{2}{c}{$\bar{B}\to\rho\mu^+\mu^-$}(Belle)& \multicolumn{2}{c|}{LHCb (untagged)}\\
		\hline
		Observable& [0.1-1] GeV$^2$& [1-6] GeV$^2$&[0.1-1] GeV$^2$& [1-6] GeV$^2$& [0.1-1] GeV$^2$& [1-6] GeV$^2$\\
		\hline
		$\left<P_1\right>$& $0\pm0.178$ & $-0.044\pm0.119$&$0\pm0.179$ & $-0.048\pm 0.125$&$0\pm0.178$ & $-0.046\pm0.122$ \\
		$\left<P_2\right>$ & $0.0772\pm0.018$& $0.073\pm0.071$& $0.071\pm 0.017$& $-0.016\pm 0.071$&$0.009\pm 0.009$& $0.045\pm0.060$\\
		$\left<P_3\right>$& $0\pm0$& $0\pm0.001$ & $0\pm0.001$& $0.001\pm-0.004$& $0\pm0$ & $0\pm0.001$\\
		$\left<P_4^{\prime}\right>$& $-0.501\pm0.106$& $0.538\pm0.169$&$-0.501\pm0.104$& $0.597\pm0.174$& $-0.500\pm0.094$ & $0.567\pm0.151$\\
		$\left<P_5^{\prime}\right>$ & $0.455\pm0.095$& $-0.215\pm0.099$& $0.368\pm0.079$& $-0.308\pm0.100$ & $0.058\pm0.011$ & $0.043\pm0.009$ \\
		$\left<P_6^{\prime}\right>$ & $-0.0136\pm0.003$& $-0.023\pm0.005$& $-0.078\pm0.015$& $-0.061\pm0.014$ & $-0.045\pm0.009$ & $-0.042\pm0.009$	\\
		$\left<P_8^{\prime}\right>$ & $0.006\pm0.001$ & $0.010\pm0.002$ & $0.019\pm0.004$ & $0.002\pm0.004$ & $0.012\pm0.003$ & $0.014\pm0.003$\\
		$\left<R_{\rho}\right>$& $0.958\pm0.181$& $1.128\pm0.263$& $0.961\pm0.174$& $1.124\pm 0.265$& $0.956\pm0.165$ & $1.116\pm0.245$ \\ 
			$\left<BR\right>\times 10^{9}$& $3.688\pm0.515$ & $7.052\pm 1.23$ & $3.282\pm0.451$ & $6.892 \pm 1.211$& $3.485\pm 0.483$ & $6.972\pm 1.221$\\
		$\left<A_{FB}\right>$ & $-0.053 \pm 0.005$& $-0.027\pm0.019$& $-0.046\pm 0.005$& $0.004\pm 0.018$ & $-0.006\pm0.001$ & $-0.016\pm0.003$\\
		$\left<F_L\right>$& $0.259\pm0.064$& $0.734\pm0.220$ & $0.298\pm0.073$& $0.749\pm0.220$ & $0.278\pm0.068$ & $0.742\pm0.218$\\ \hline
	\end{tabular}
	\caption{\label{rhowoCorrec}Binned values of observables for the process $B\to\rho\mu^+\mu^-$ to be measured at Belle, $\bar{B}\to\rho\mu^+\mu^-$ to be measured at Belle for tagged events and at LHCb for untagged events are given.}
\end{table*}

As we mentioned above, due to $\xi_u/\xi_t$ term in the present case, which is practically negligible in the case of $b\to s $ transition, the observable for the mode and the CP conjugated mode show clear differences and hence is a clear sign of CP violation. A precise measurement would determine whether the amount of CP violation is in conformity with the CKM picture or there are extra phases present. The observable $P_6^{\prime}$ is of particular interest in this regard as it is proportional to an imaginary part of Wilson coefficients. It can be noted that its value in low $q^2$ is significantly different for CP-conjugate modes, giving large value of CP asymmetry. \newline

From the Table \ref{KstarwoCorrec}, it is easy to note that different choices of form factors yield values for FFI observables that are reasonably close to each other while for FFD observables, like branching ratio, the impact is significant and there is a larger spread in the predictions. 
Comparing $\bar{B}^0\to\rho^0\mu^+\mu^-$ with $\bar{B}_s\to K^{\ast}\mu^+\mu^-$, effects of strange quark versus up/down quark is apparent in many observables. Along with the observables discussed in the previous section, we report branching ratio for $B^0\to\rho^0\mu^+\mu^-$ and $B_s\to\bar{K}^{\ast}\mu^+\mu^-$ over the full kinematically allowed range. For $B^0\to\rho^0\mu^+\mu^-$, time-integrated branching ratio within SM is found out to be, 

\begin{eqnarray}\label{BRrho}
	\left<\mathcal{BR}(B\to\rho\mu^+\mu^-)\right>^{\text{Belle}}&=&(4.131\pm 0.679)\times 10^{-8}\nn\\
	\left<\mathcal{BR}(\bar{B}\to\rho\mu^+\mu^-)\right>^{\text{Belle}}&=&(4.198\pm 0.678)\times 10^{-8}\nn\\
	\left<BR\right>^{\text{LHCb}}&=&(4.164\pm 0.678)\times 10^{-8}
\end{eqnarray}

From the definition of observables given in Appendix \ref{appendixA} it is clear that the measurable quantity is actually time-integrated branching ratio normalized by decay rate. For results given in Eq. (\ref{BRrho}), we have taken the mean value of decay rate to be $\Gamma_{B^0}=6.579\times 10^{11} s^{-1}$ \cite{PDG}. Thus, the actual observable $\left<d\Gamma/dq^2\right>^{\text{Belle}}$ defined in section \ref{appendixA} is obtained by multiplying the results in Eq. (\ref{BRrho}) by $\Gamma_{B^0}^{-1}$. Branching ratio of $\bar{B}_s\to K^{\ast}\mu^+\mu^-$ in SM using BSZ form factors based on LCSR calculation is,   
\begin{eqnarray}\label{BRkstar}
	\mathcal{BR}(B_s\to \bar{K}^{\ast}\mu^+\mu^-)&=&(2.849\pm 0.719 )\times 10^{-8}\nn\\
	\mathcal{BR}(\bar{B}_s\to K^{\ast}\mu^+\mu^-)&=&(2.897\pm 0.732 )\times 10^{-8}
\end{eqnarray} 
For branching ratio in full kinematic range, form factors based on LCSR are not much reliable as they are valid in low-$q^2$ region only while the kinematic range extends upto $\sim 20$ GeV$^2$ ($(M_{B_s}-M_{K^{\ast}})^2$). Hence, we also give below values of branching ratio using form factors obtained from combined fits of lattice and LCSRs results.

\begin{eqnarray}
	\mathcal{BR}(B_s\to \bar{K}^{\ast}\mu^+\mu^-)&=&(3.356\pm 0.814 )\times 10^{-8}\nn\\
	\mathcal{BR}(\bar{B}_s\to K^{\ast}\mu^+\mu^-)&=&(3.419\pm 0.827 )\times 10^{-8}
\end{eqnarray}

While finalizing this manuscript, LHCb collaboration reported a preliminary result $\mathcal{BR}(B_s\to \bar{K}^{\ast}\mu^+\mu^-)=(3.0\pm1.0\pm0.2\pm0.3)\times 10^{-8}$ \cite{LHCbprem} which is consistent with the SM prediction given in this paper.\newline

\begin{figure*}[b]
		\begin{tabular}{cc}
		\includegraphics[scale=0.5]{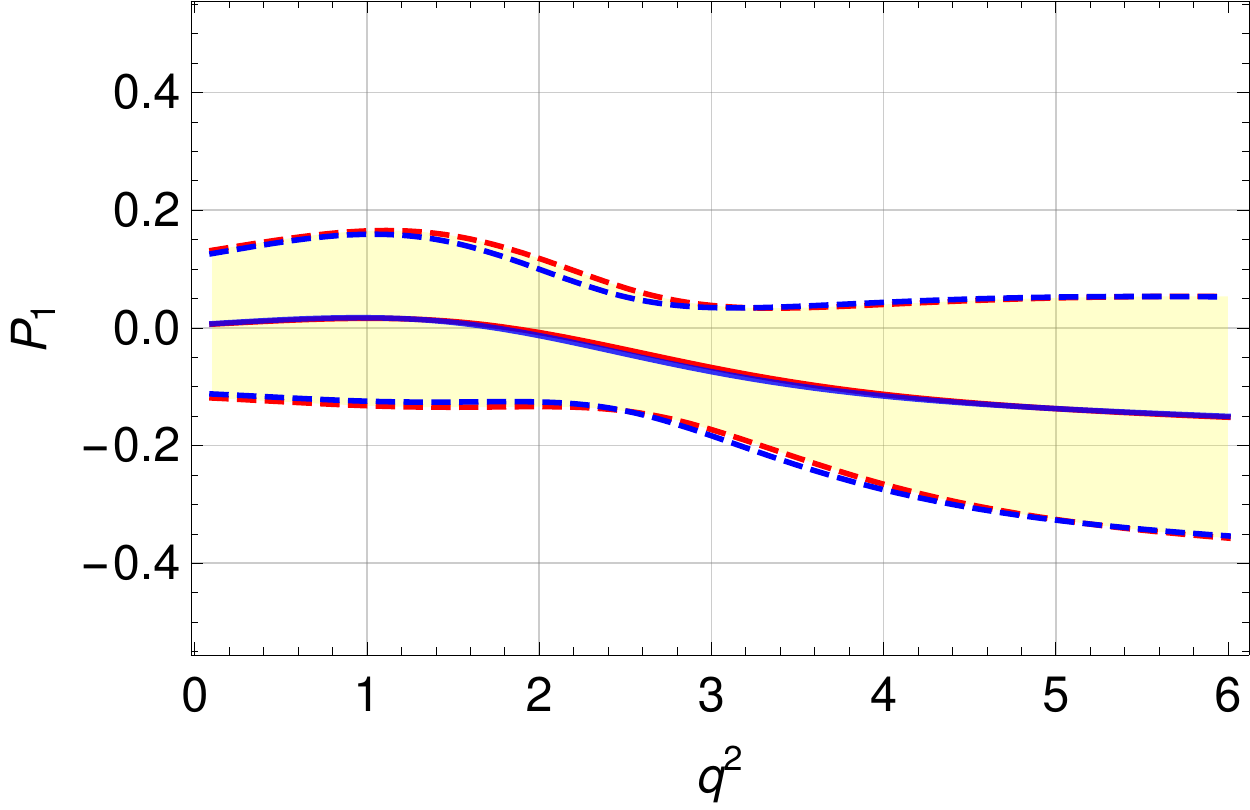}
		~~~~~	&~~~~ 
		\includegraphics[scale=0.5]{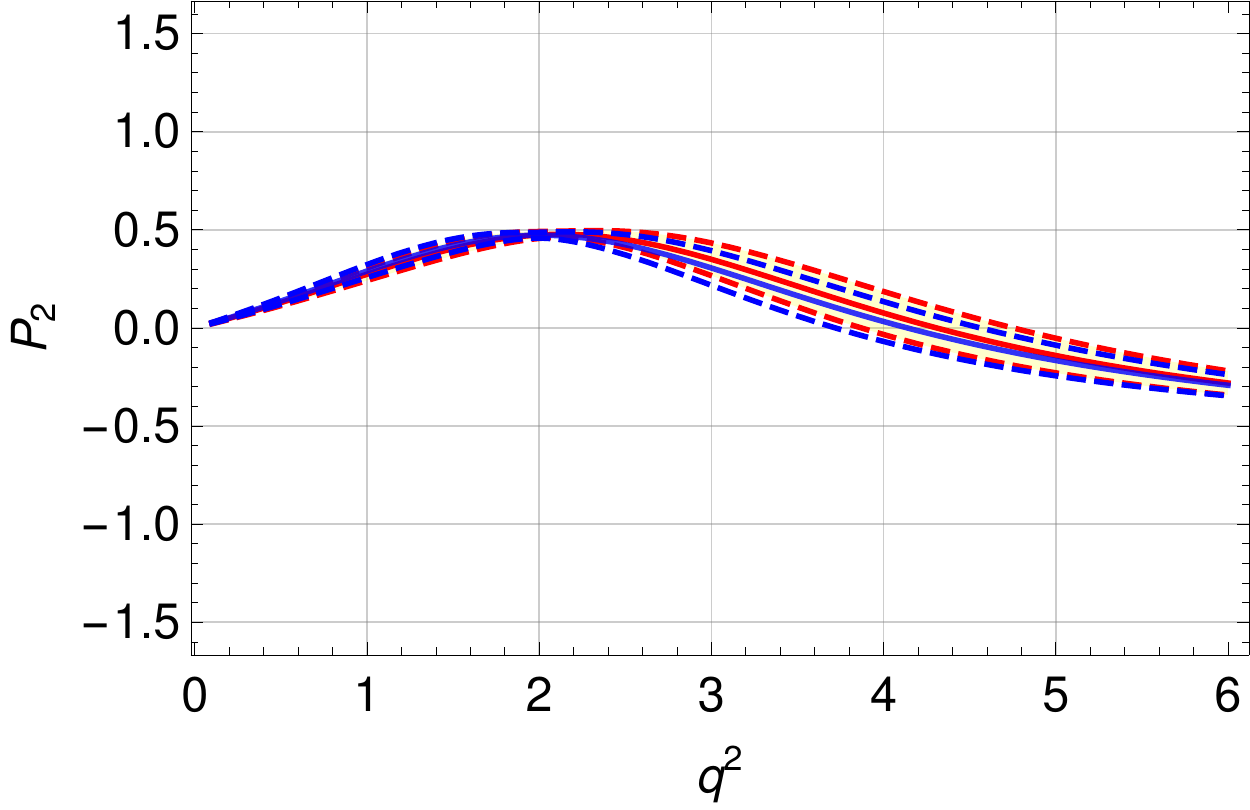}\\
		(a)& (b)\\
		\includegraphics[scale=0.5]{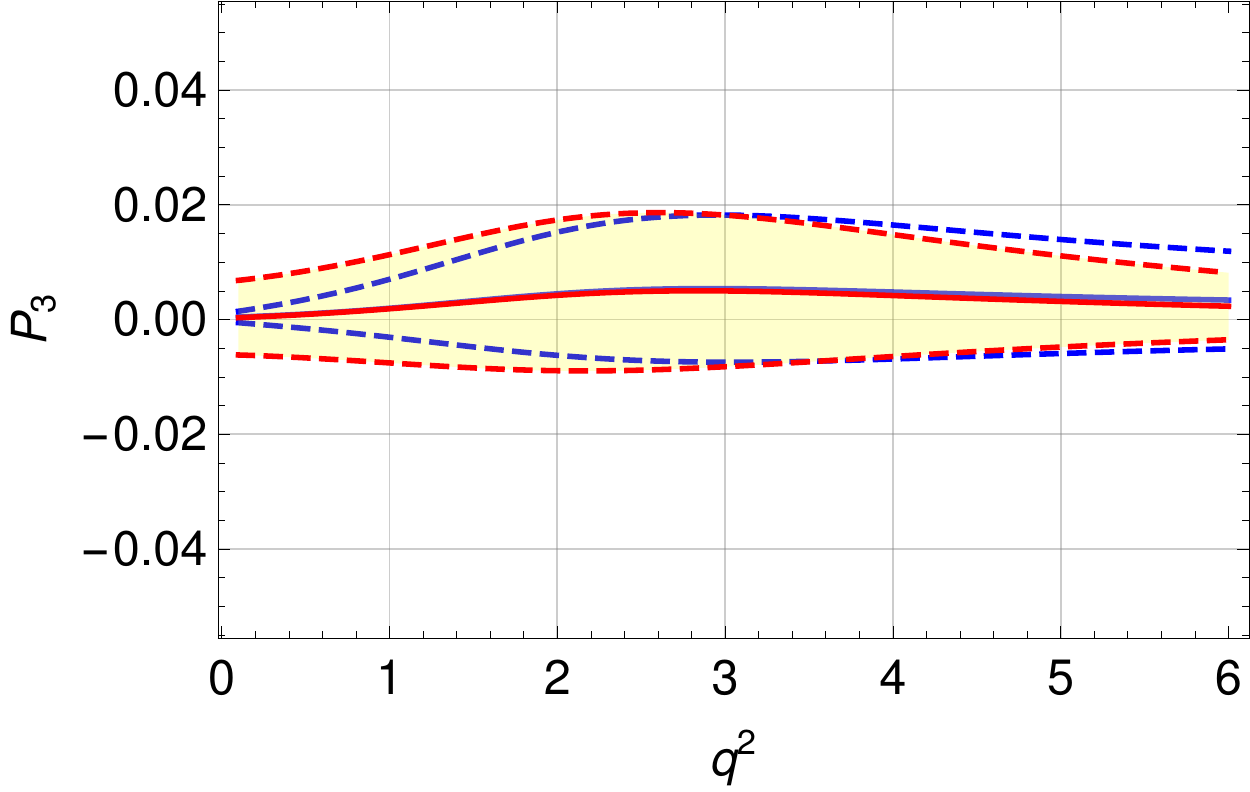}
		~~~~~	&~~~~ 
		\includegraphics[scale=0.5]{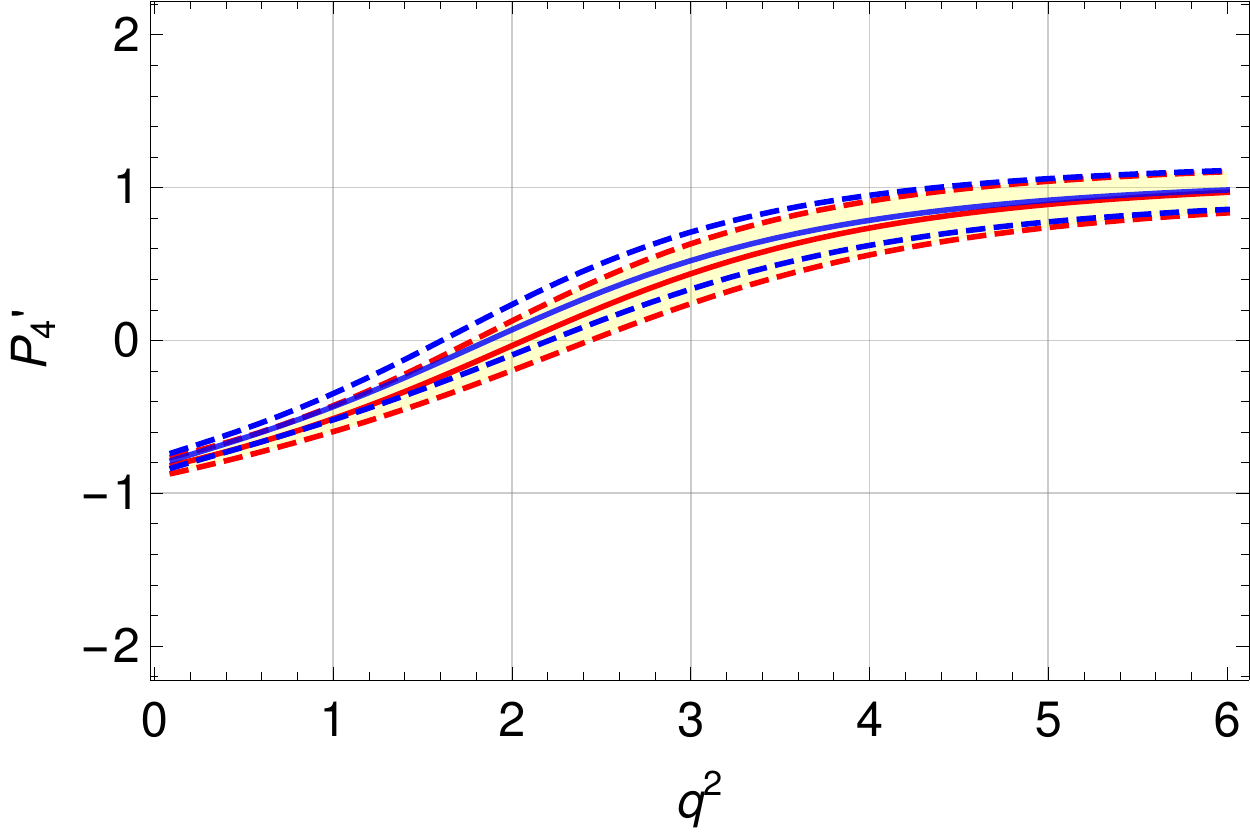}\\
		(c)& (d)\\
		\includegraphics[scale=0.5]{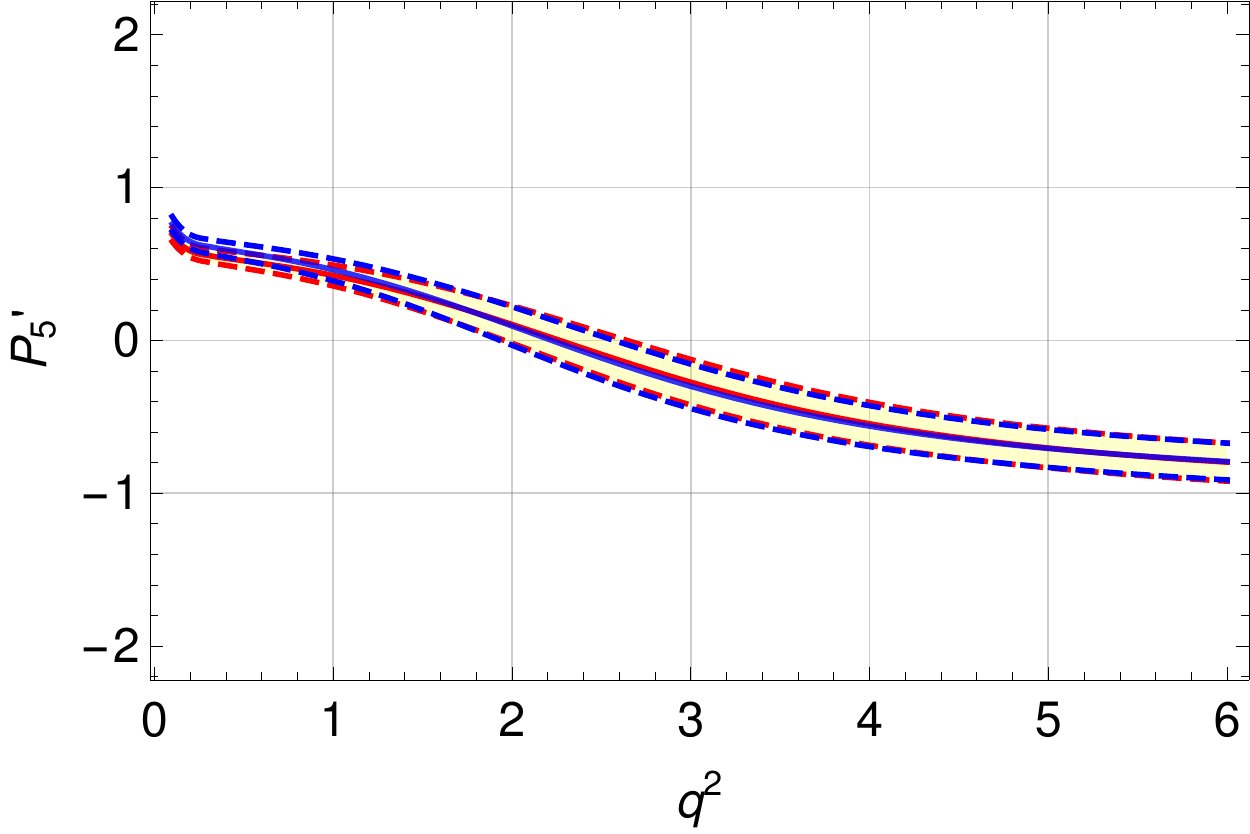}
		~~~~~	&~~~~ 
		\includegraphics[scale=0.5]{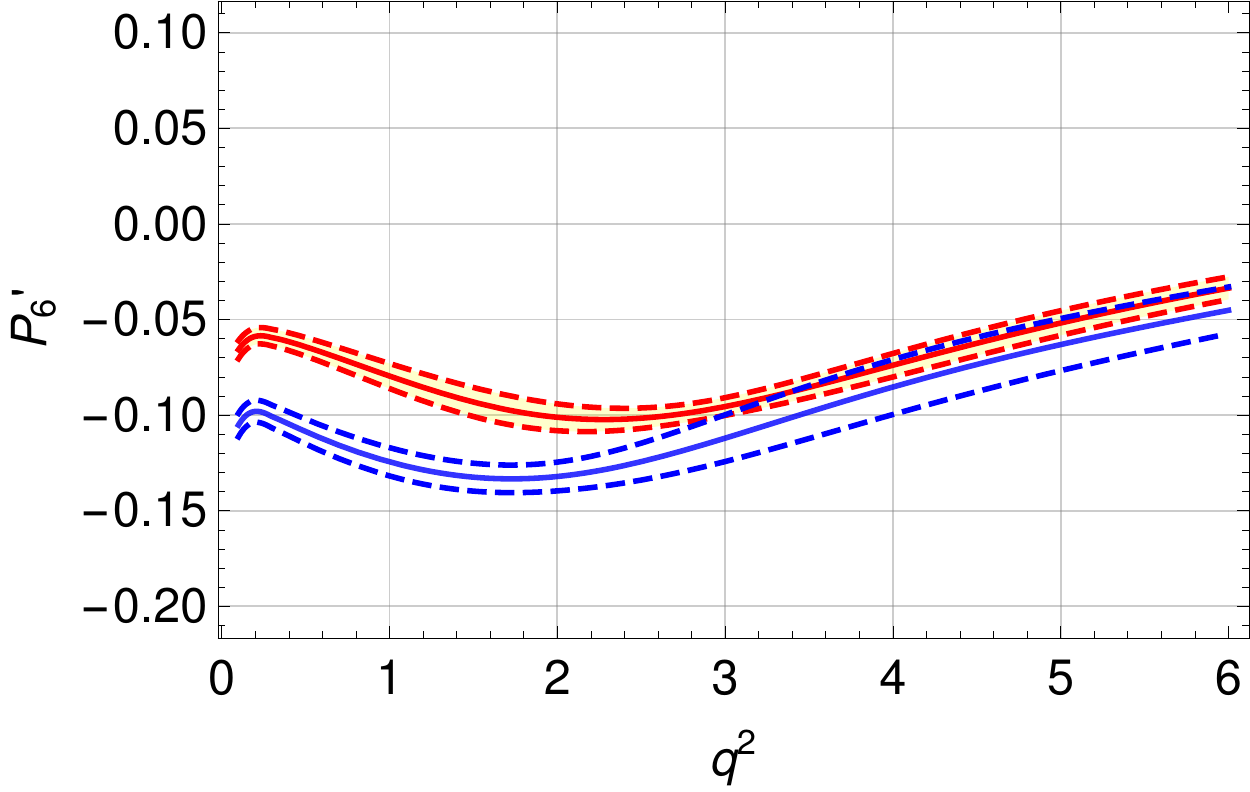}\\
		(e)& (f)	\\
		\includegraphics[scale=0.5]{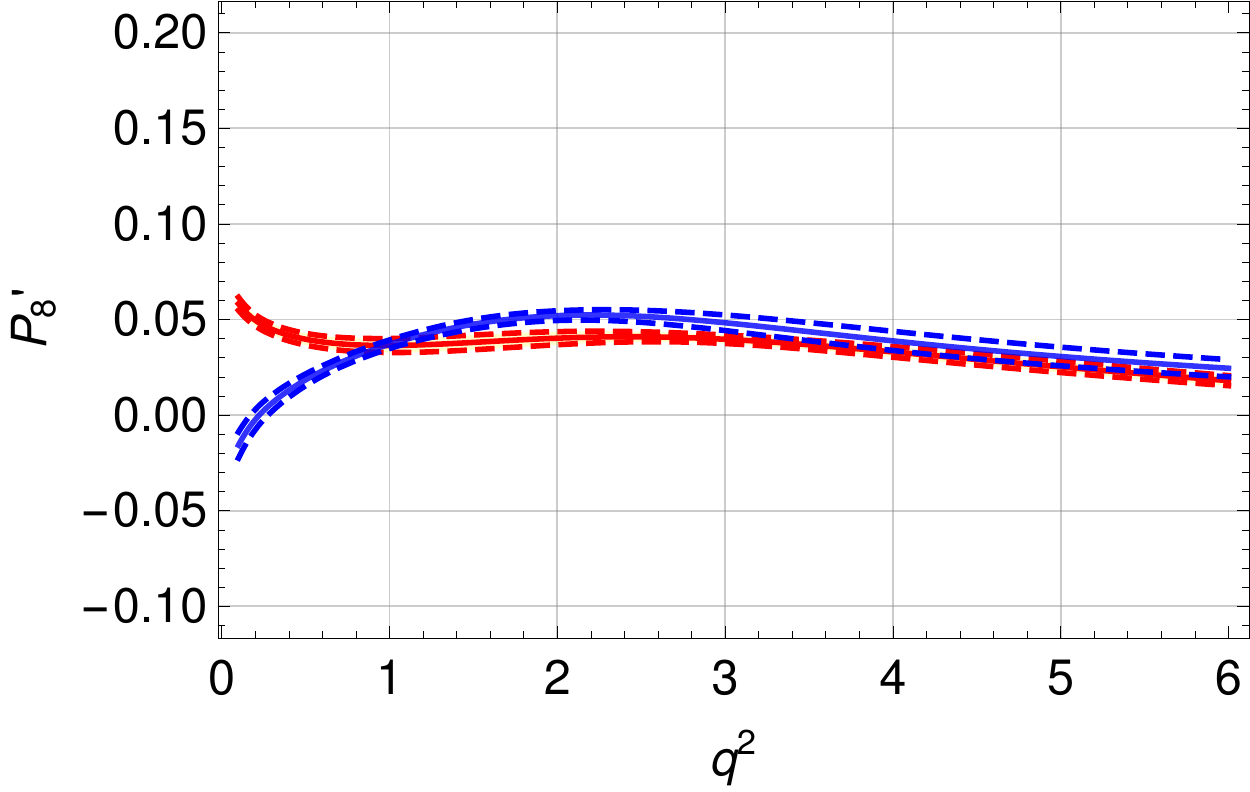} ~~~~& ~~~~
		\includegraphics[scale=0.5]{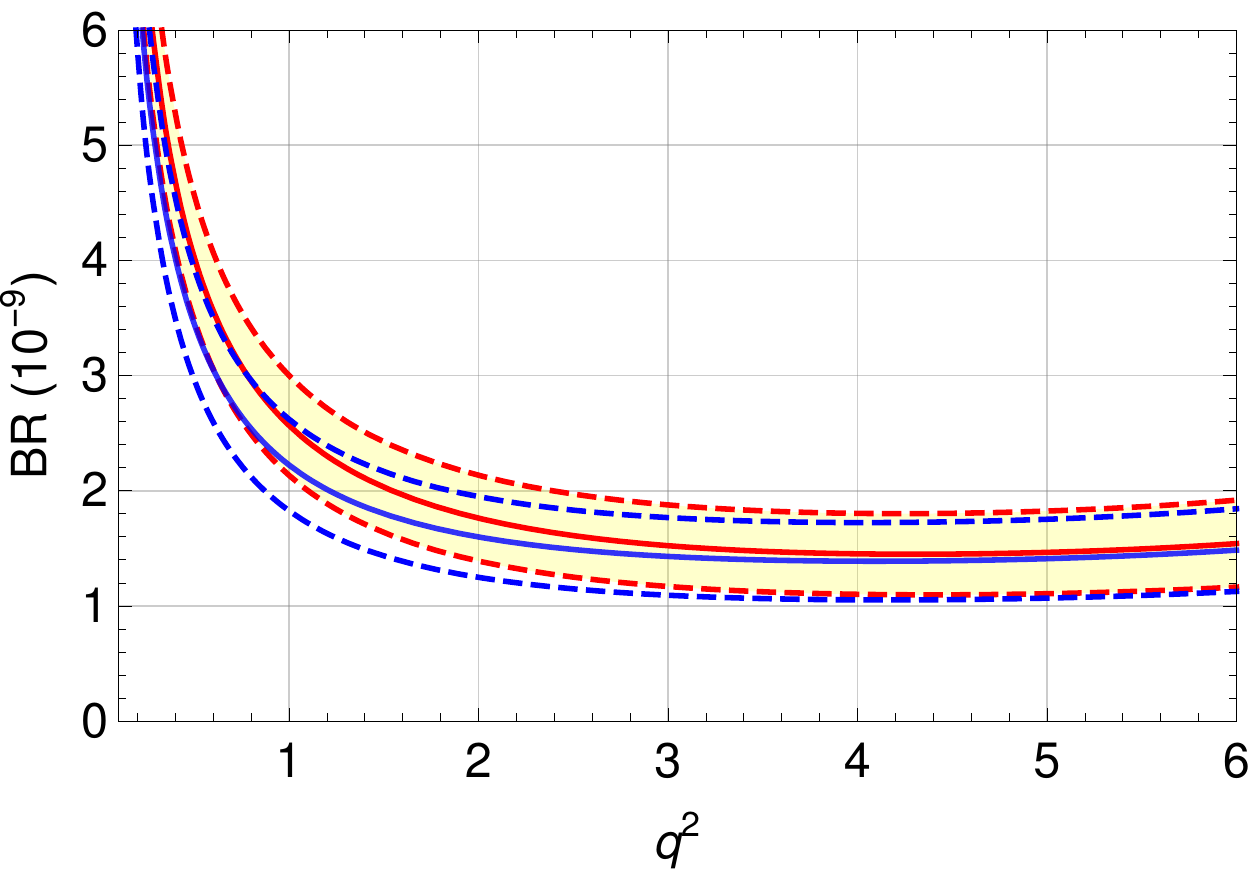}\\
		(g)& (h)\\
		\includegraphics[scale=0.5]{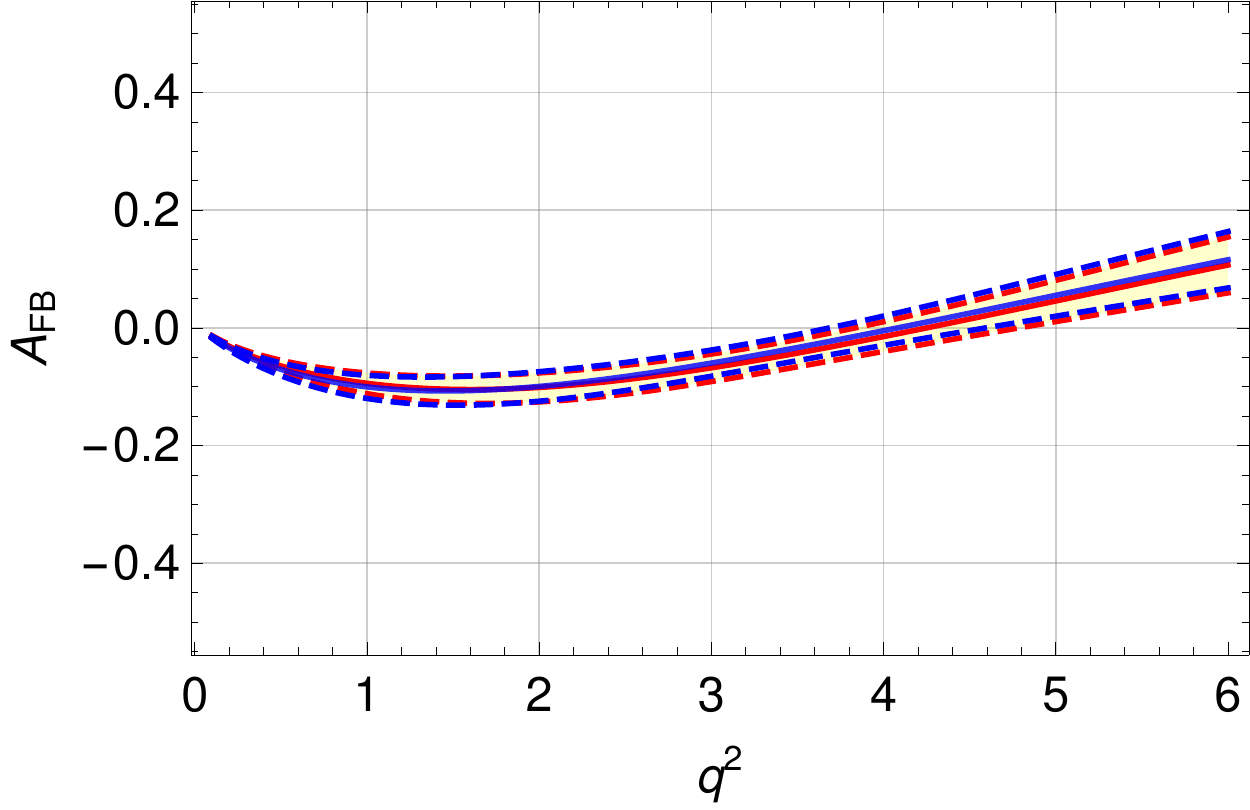}
		~~~~ & ~~~~
		\includegraphics[scale=0.5]{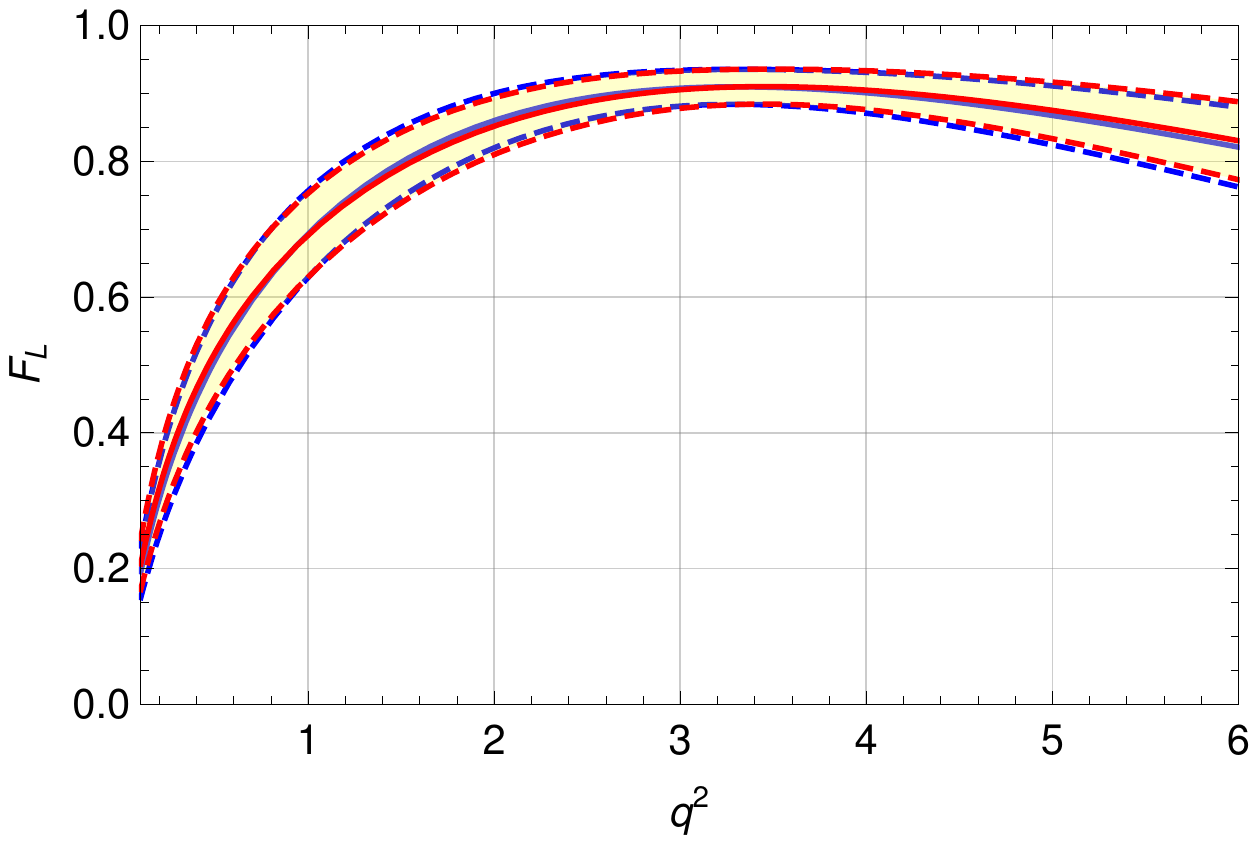}\\
		(i) & (j)
	\end{tabular}
	\caption{\label{kstarplots}Observables as functions of $q^2$. Red solid curve shows the mean value of observable for $B_s\to \bar{K^{\ast}}\mu^+\mu^-$, Blue solid curve show mean value of observables for $\bar{B}_s\to K^{\ast}\mu^+\mu^-$. Blue dashed and Red dashed curve show uncertainty in the values. These plots are obtained using BSZ2 form factors. The mean values include the contribution of non-factorizable corrections. The uncertainty in the bands is due to errors in determination of form factors only.}
\end{figure*}

\begin{figure*}[b]
	\begin{tabular}{cc}
		\includegraphics[scale=0.5]{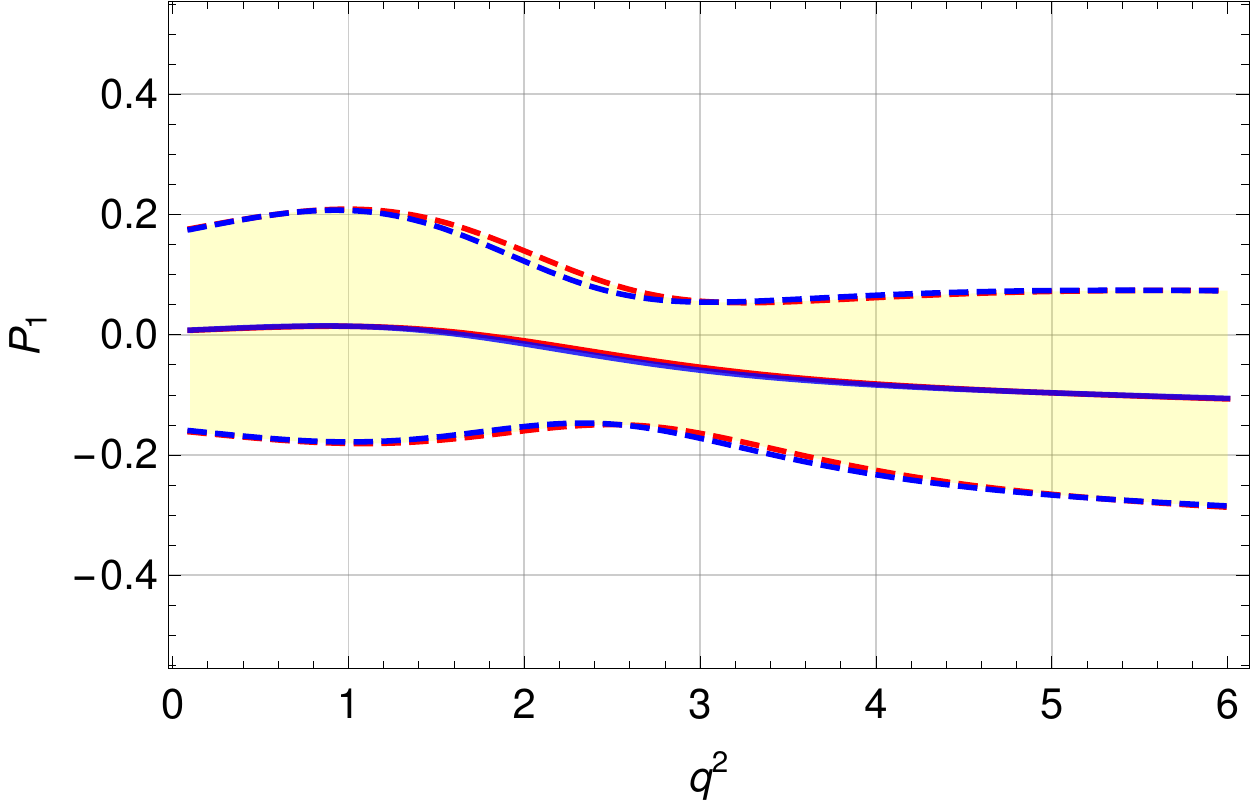}
		~~~~~	&~~~~ 
		\includegraphics[scale=0.5]{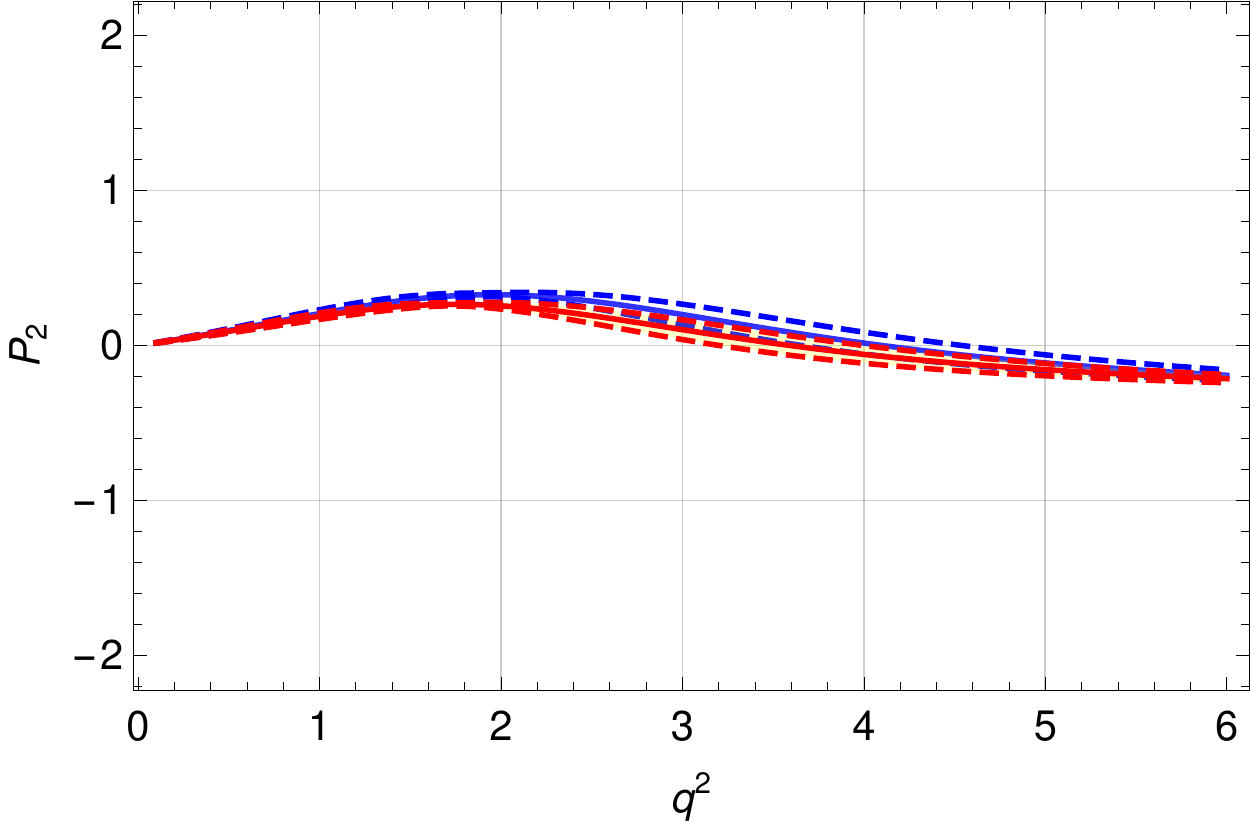}\\
		(a)& (b)\\
		\includegraphics[scale=0.5]{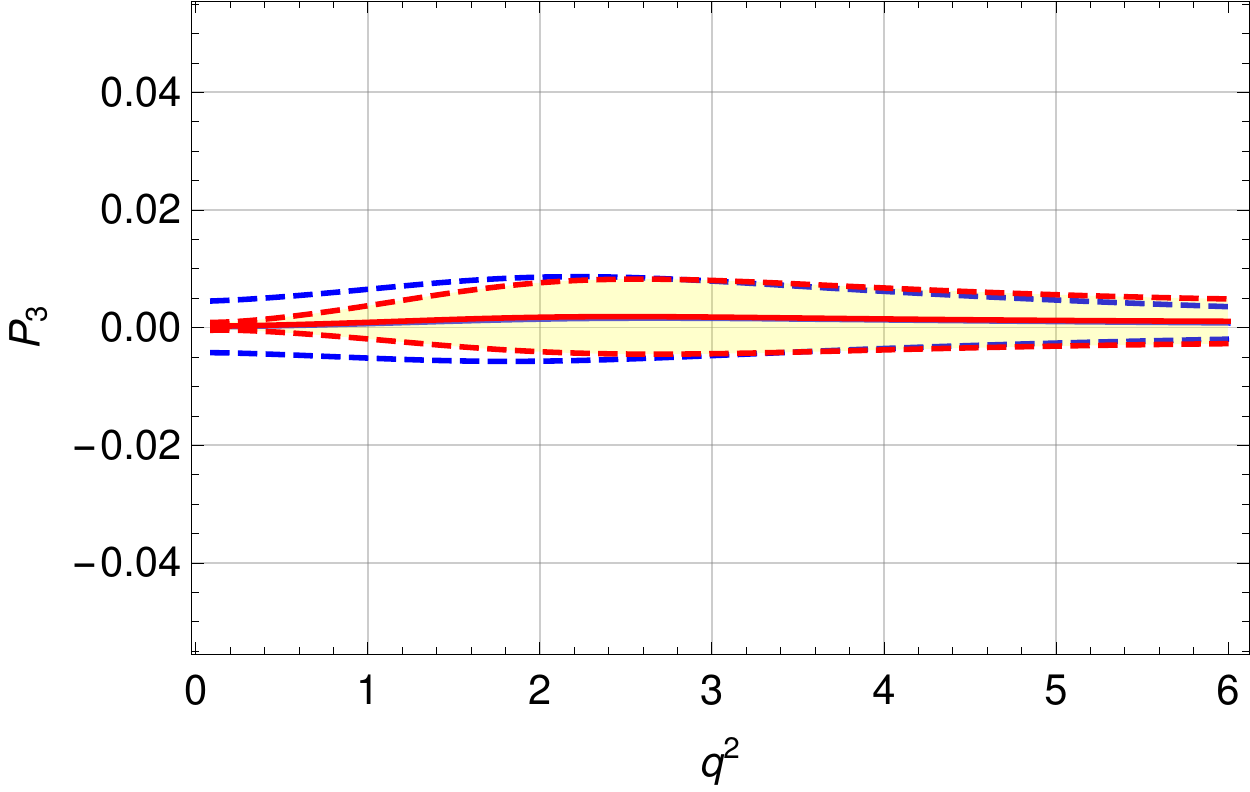}
		~~~~~	&~~~~ 
		\includegraphics[scale=0.5]{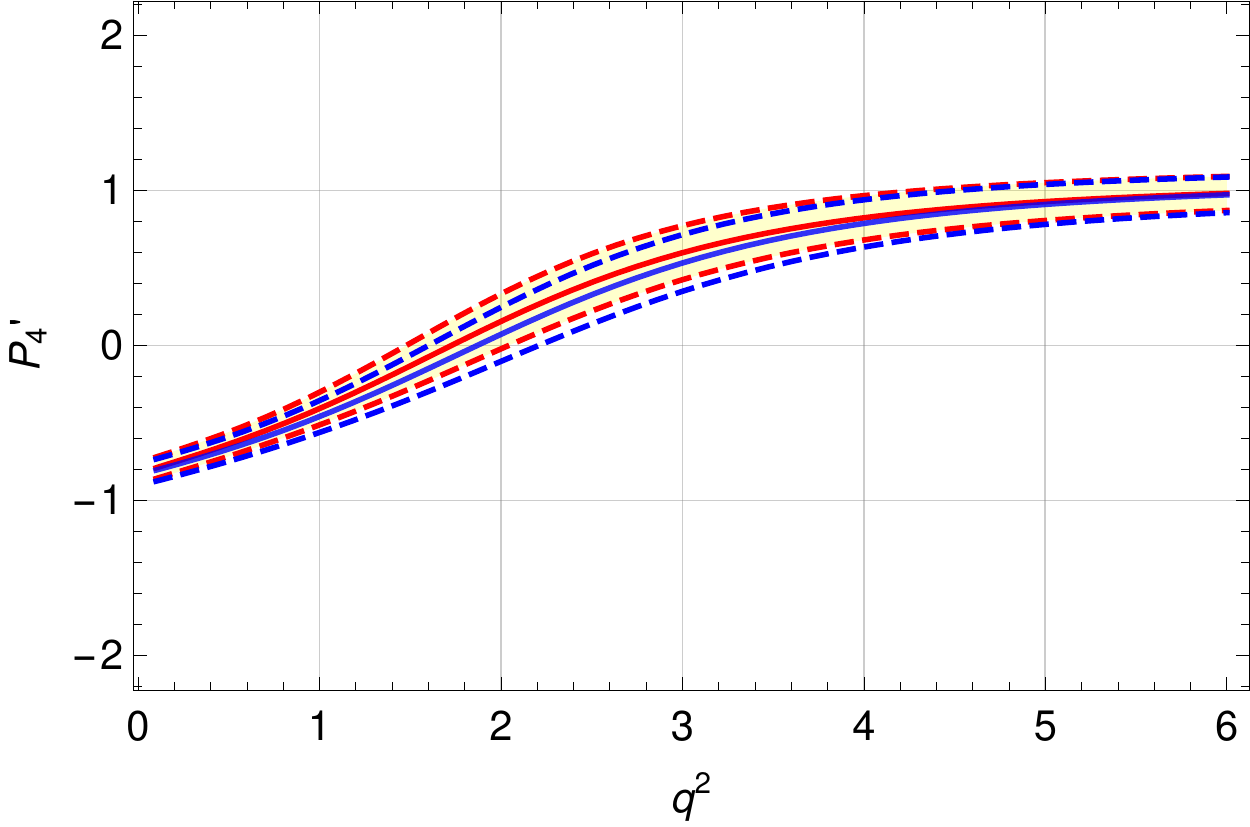}\\
		(c)& (d)\\
		\includegraphics[scale=0.5]{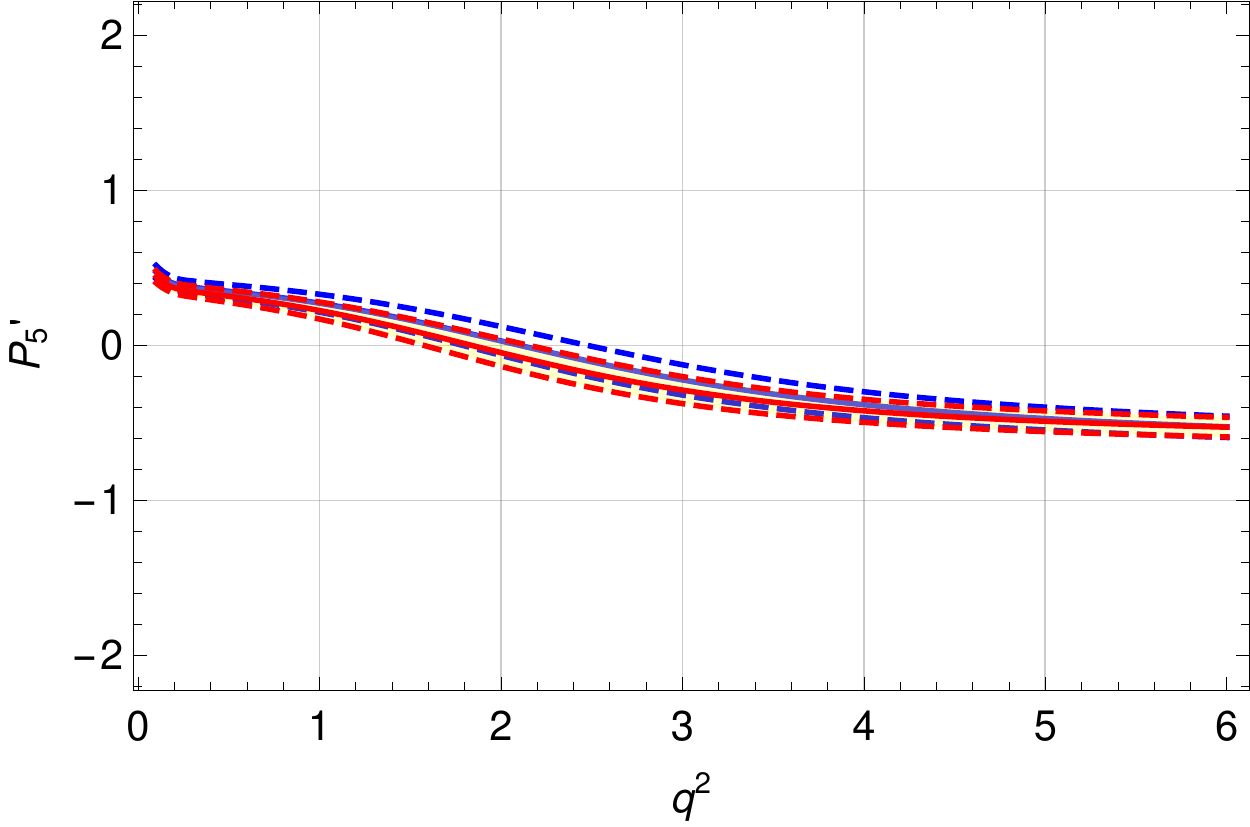}
		~~~~~	&~~~~ 
		\includegraphics[scale=0.5]{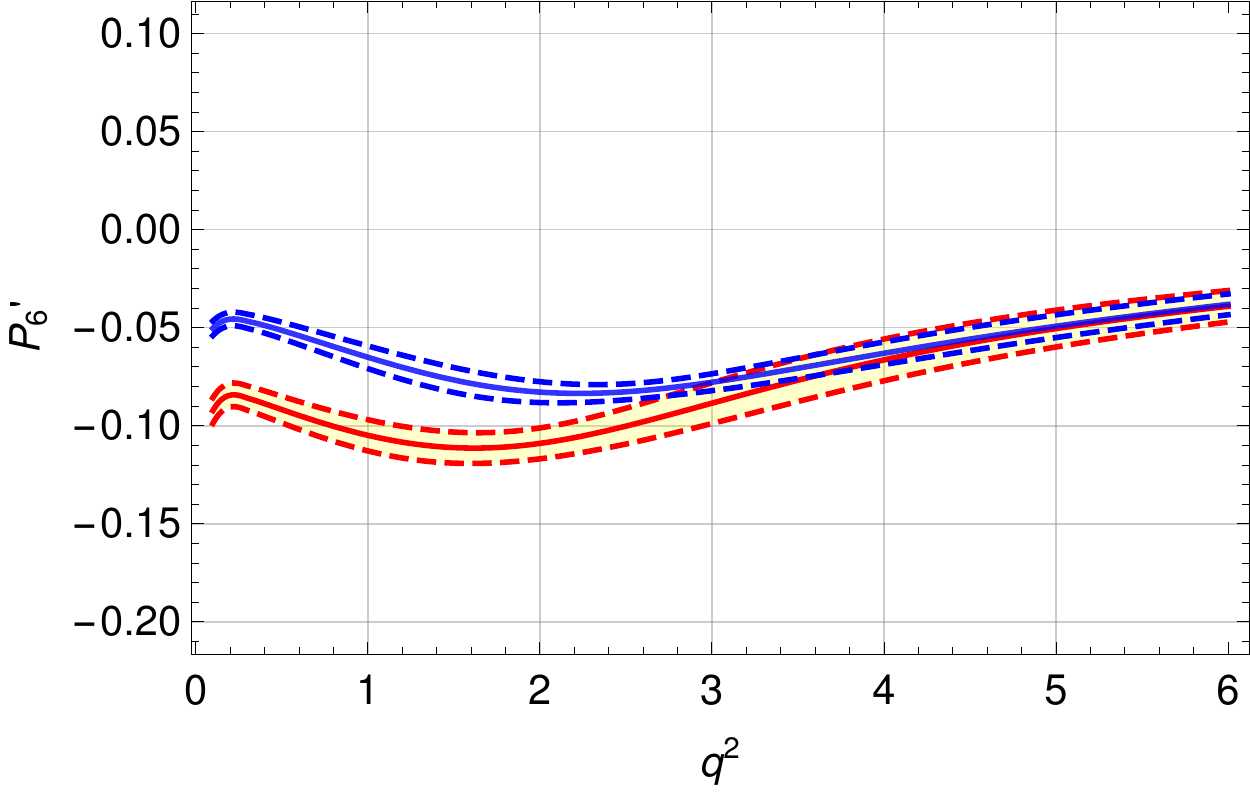}\\
		(e)& (f)	\\
	\includegraphics[scale=0.5]{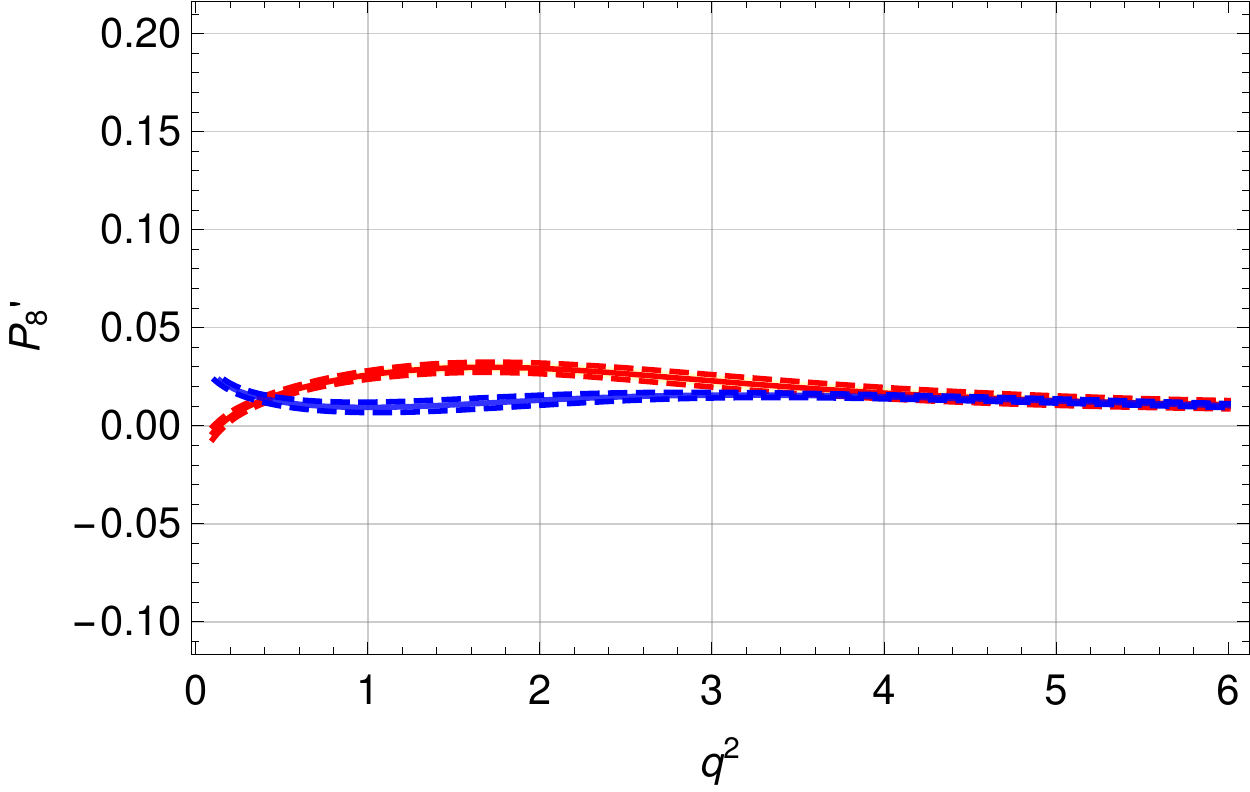} ~~~~& ~~~~
        \includegraphics[scale=0.5]{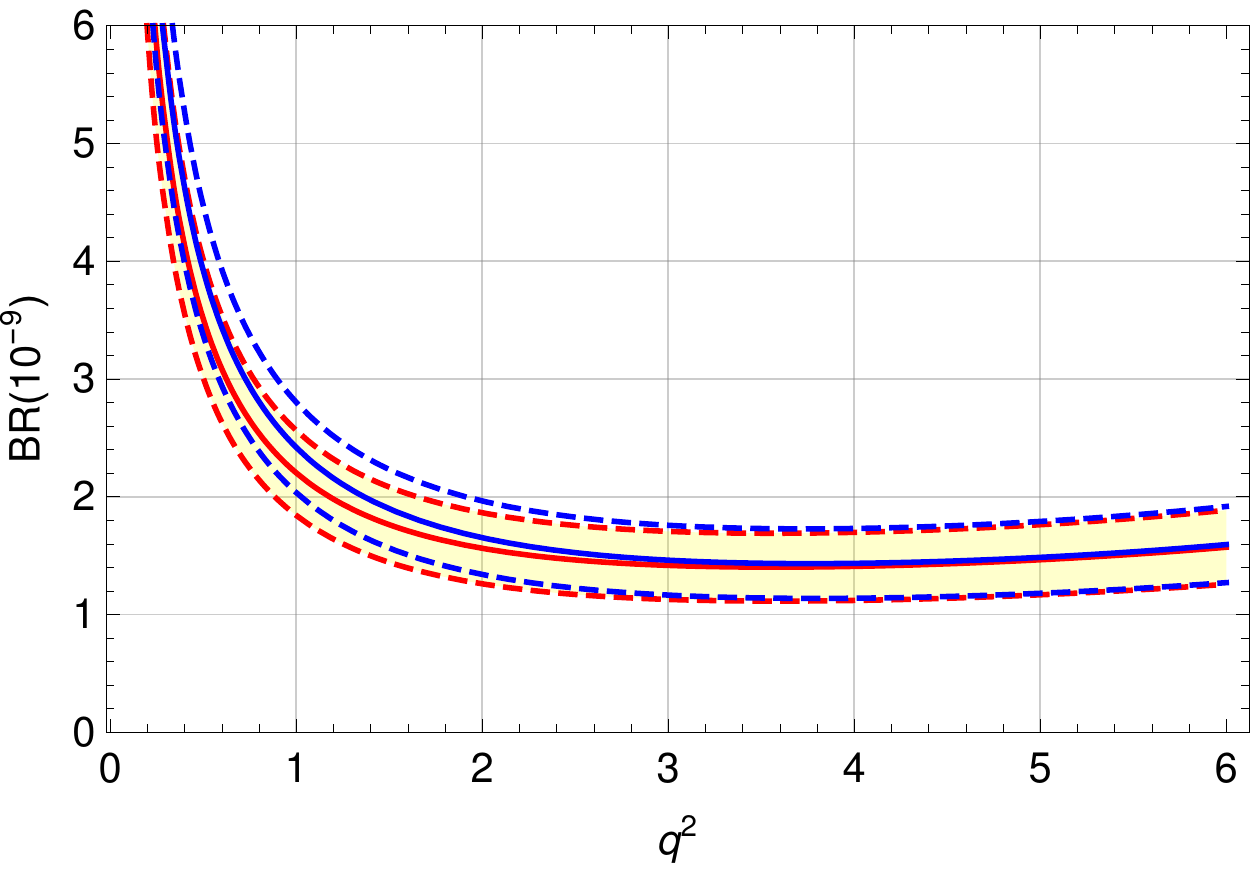}\\
        (g)& (h)\\
        \includegraphics[scale=0.5]{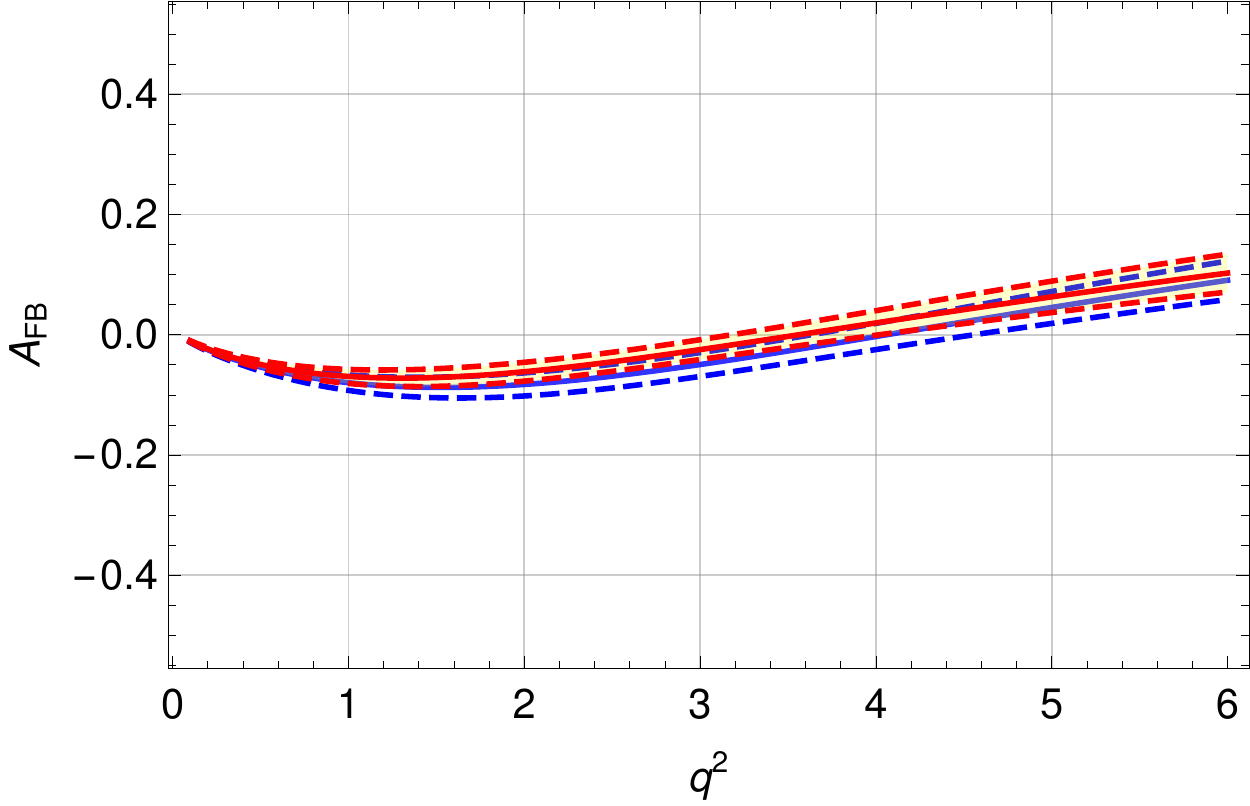}
        ~~~~ & ~~~~
        \includegraphics[scale=0.5]{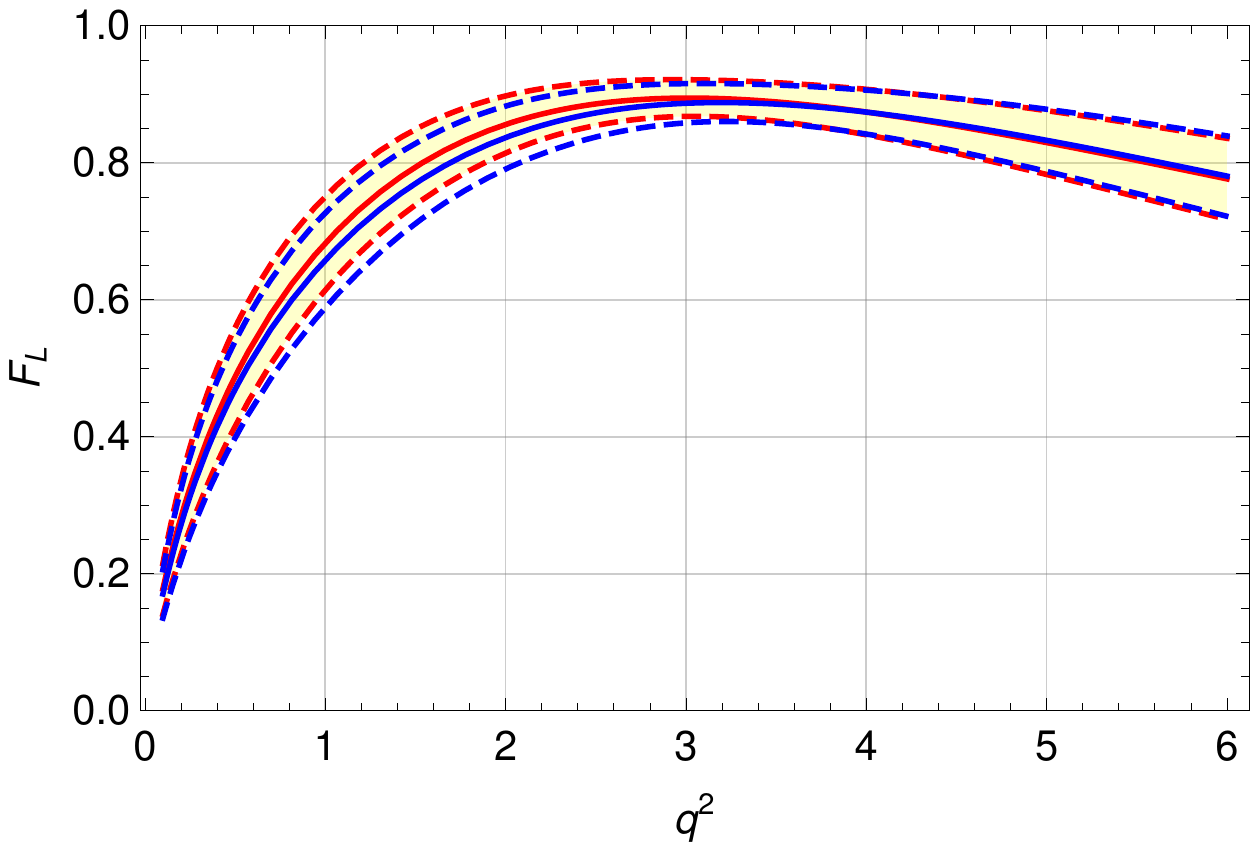}\\
        (i) & (j)
\end{tabular}
	\caption{\label{rhoBelleplots} observables as functions of $q^2$. Red solid curve shows the mean value of observable for $\bar{B}\to \rho\mu^+\mu^-$, Blue solid curve show mean value of observables for $B\to \rho \mu^+\mu^-$. Blue dashed and Red dashed curve show uncertainty in the values. These plots are obtained using BZ form factors. The mean values include the contribution of non-factorizable corrections. The uncertainty in the bands is due to errors in determination of form factors only.}
\end{figure*}

\begin{figure*}[b]
	\begin{tabular}{cc}
		\includegraphics[scale=0.5]{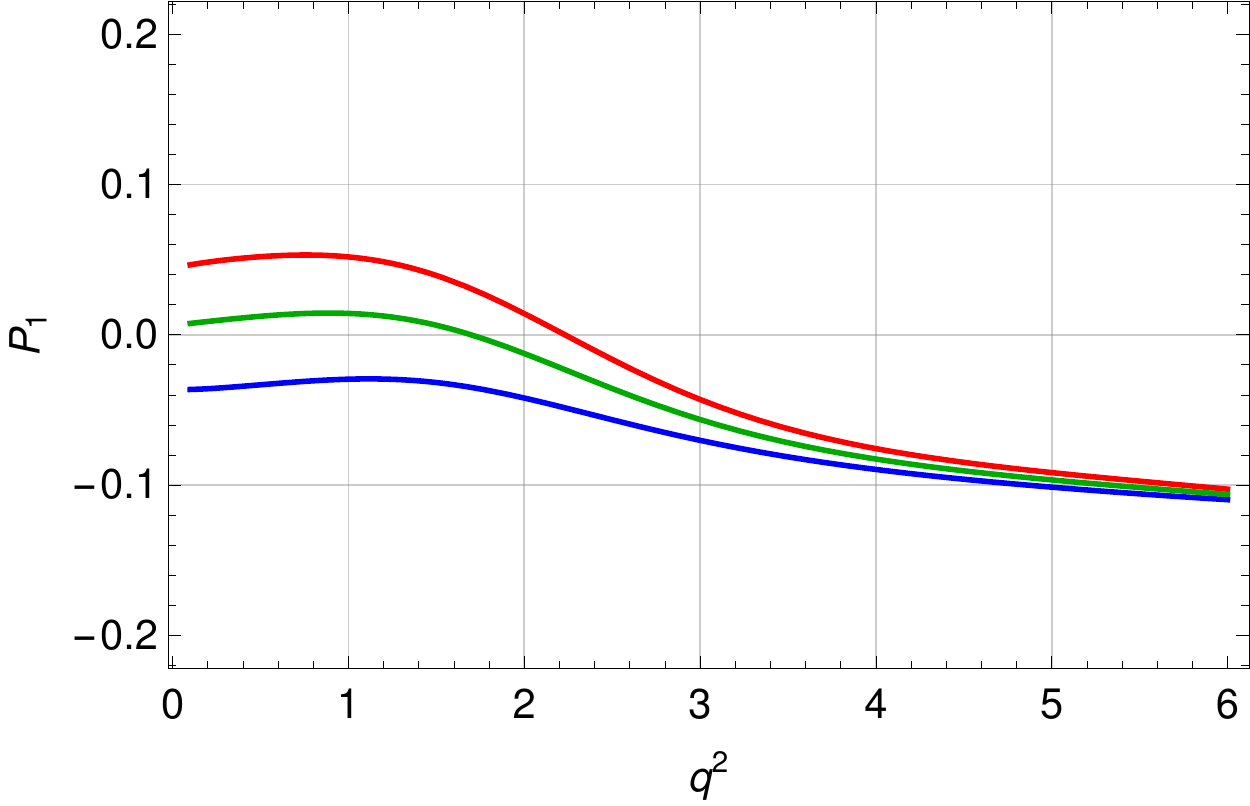}
		~~~~~	&~~~~ 
		\includegraphics[scale=0.5]{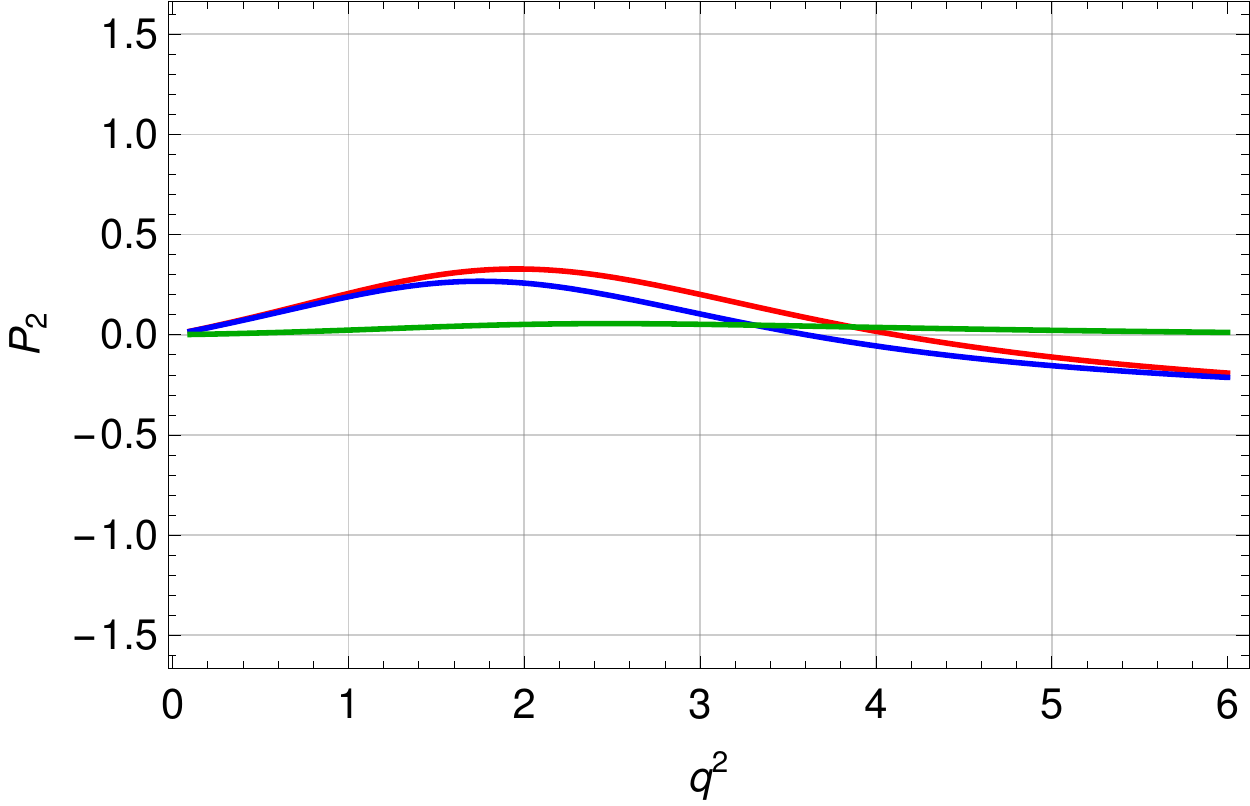}\\
		(a)& (b)\\
		\includegraphics[scale=0.5]{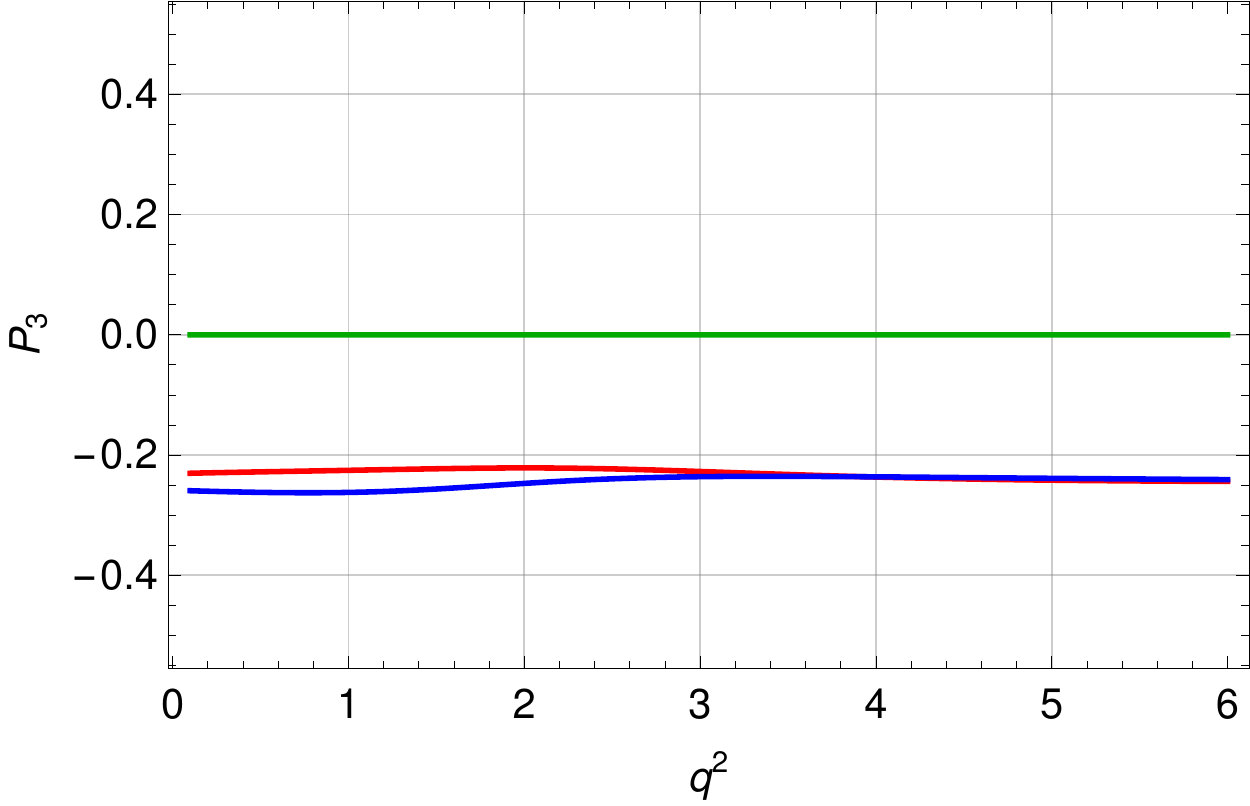}
		~~~~~	&~~~~ 
		\includegraphics[scale=0.5]{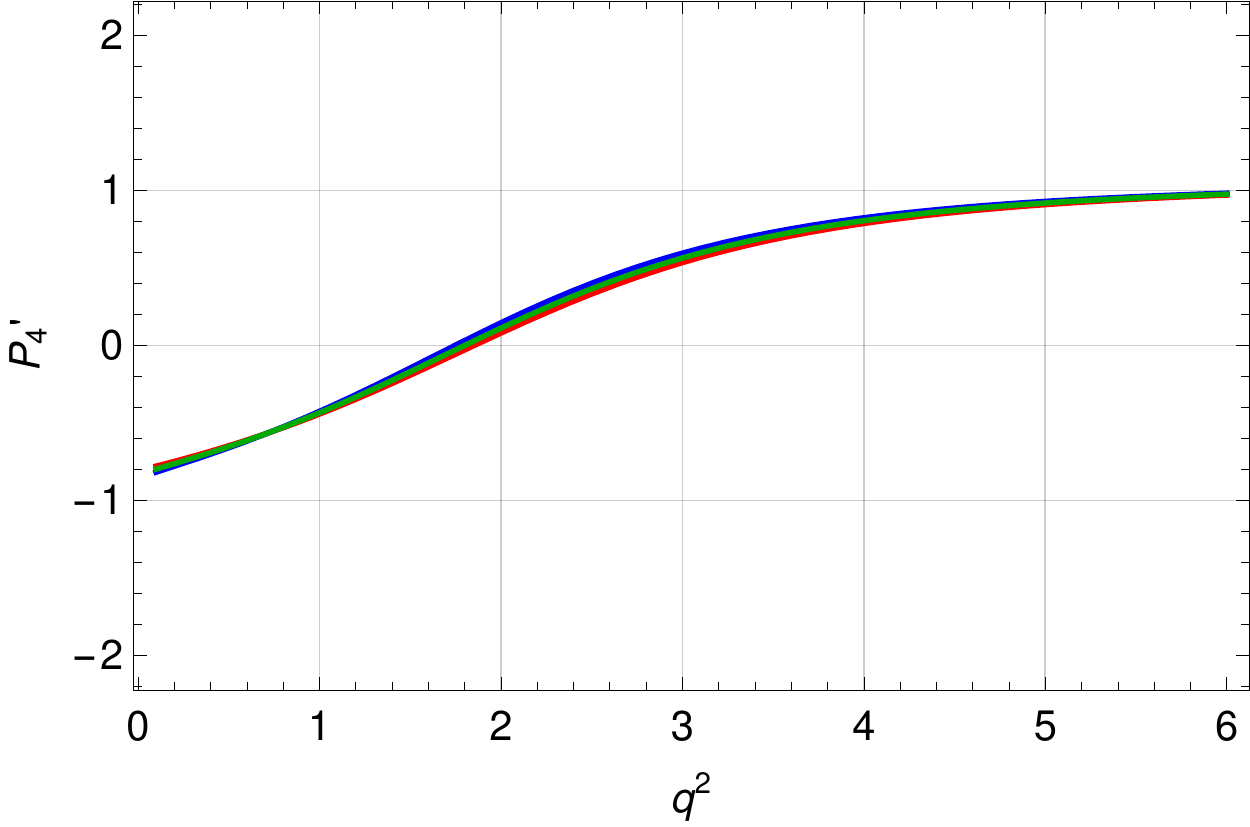}\\
		(c)& (d)\\
		\includegraphics[scale=0.5]{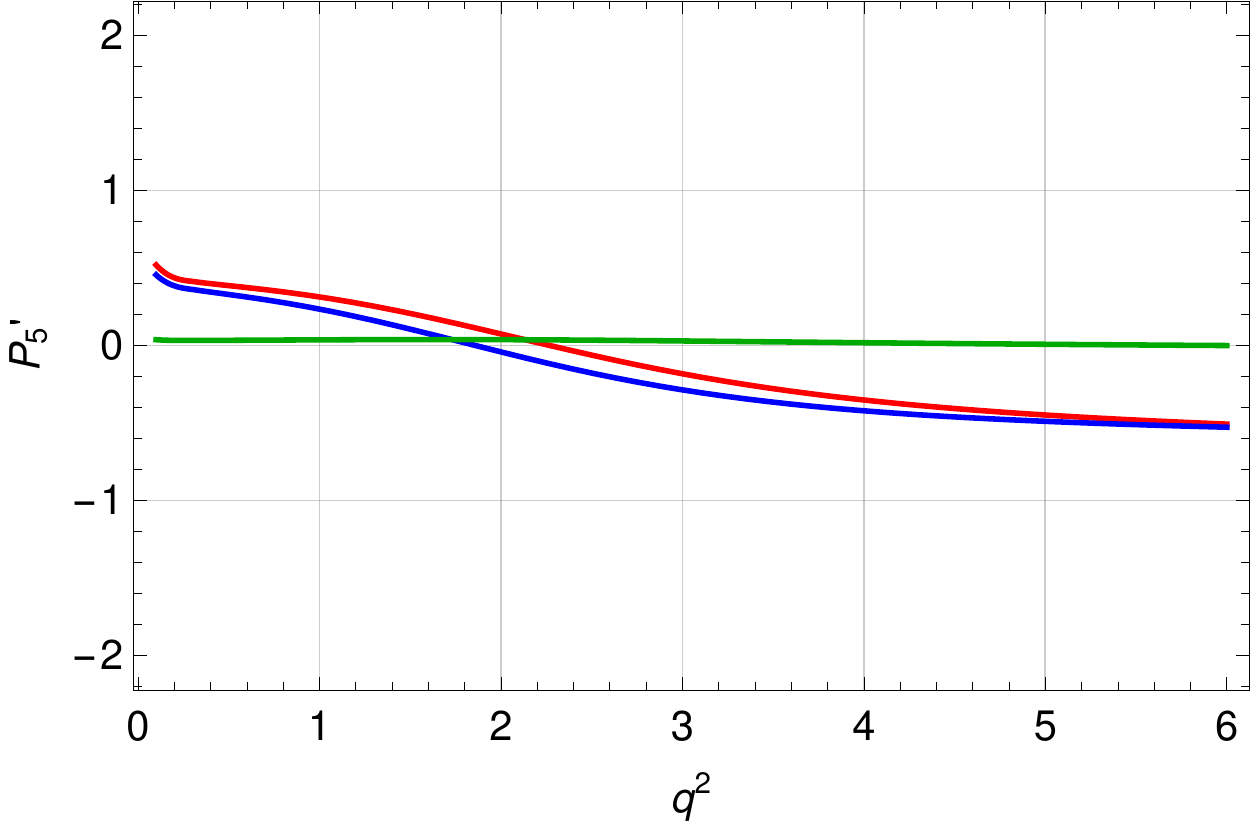}
		~~~~~	&~~~~ 
		\includegraphics[scale=0.5]{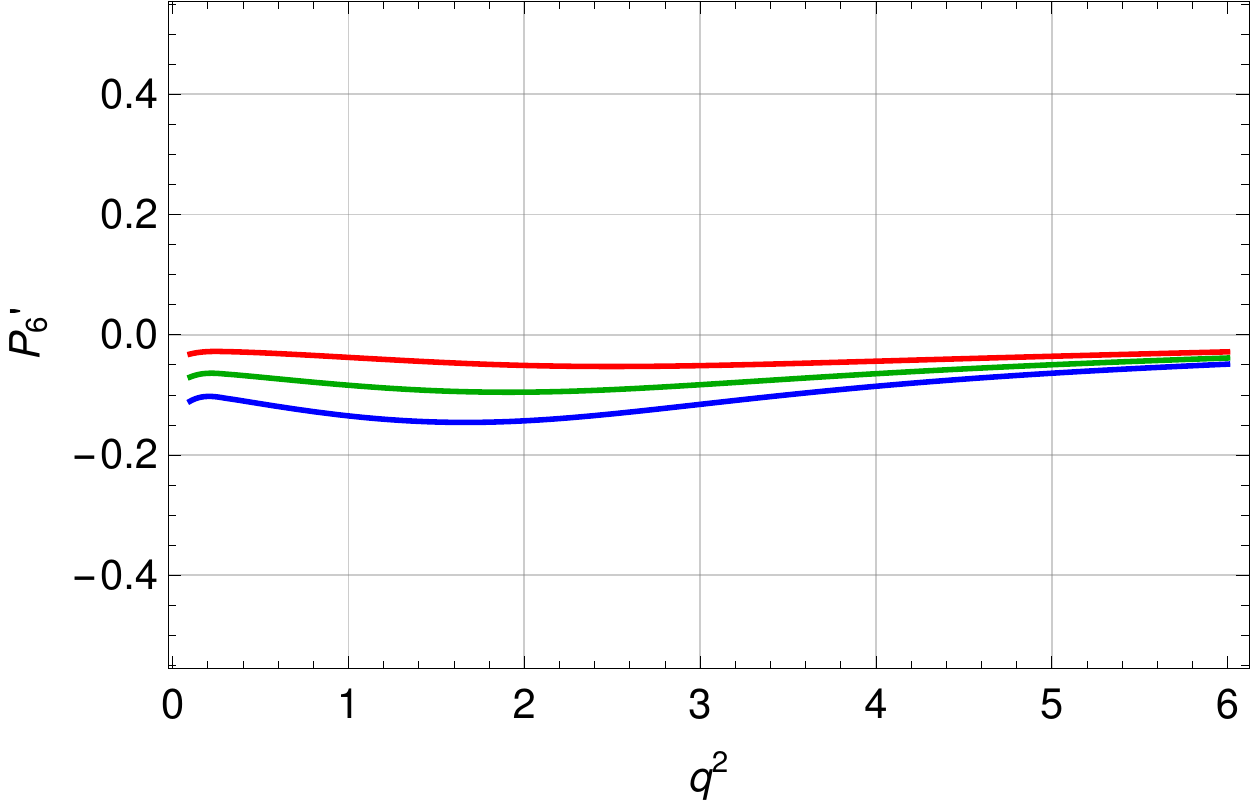}\\
		(e)& (f)	\\
		\includegraphics[scale=0.5]{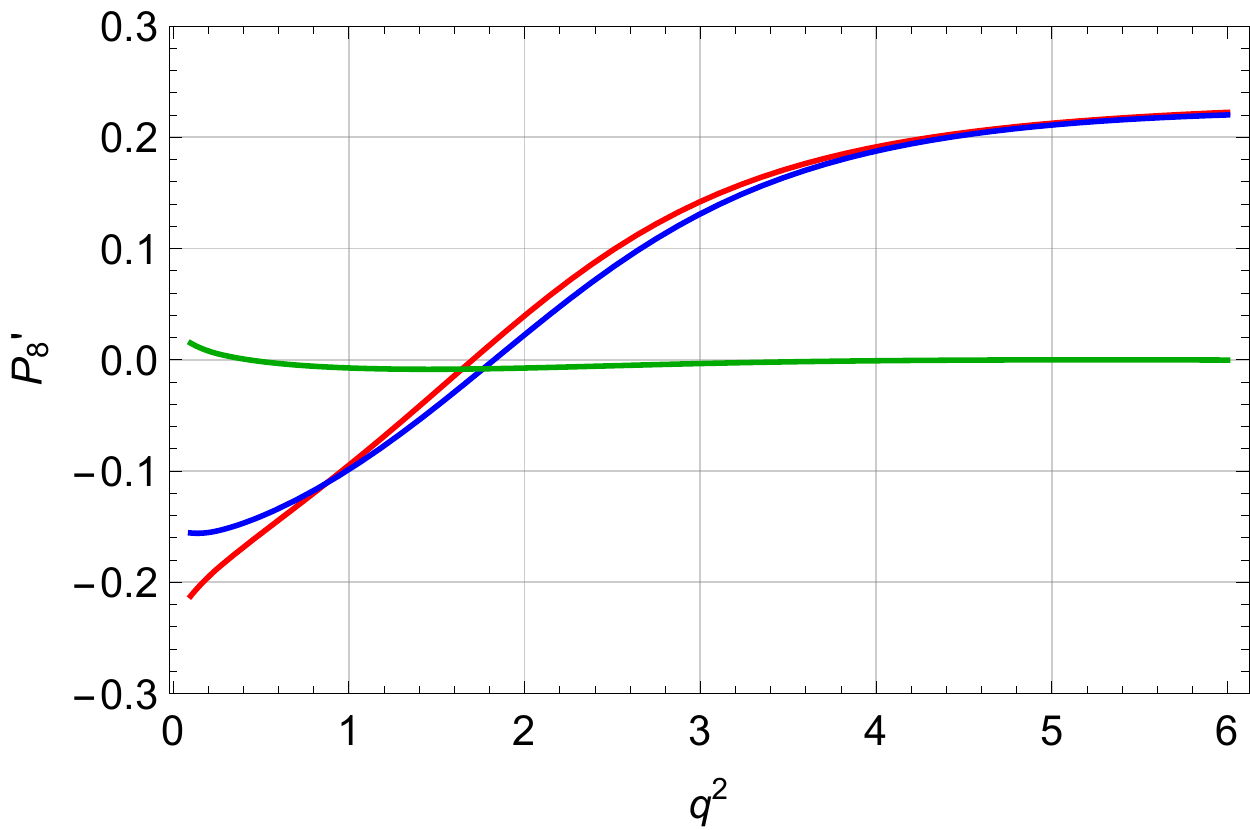} ~~~~& ~~~~
		\includegraphics[scale=0.5]{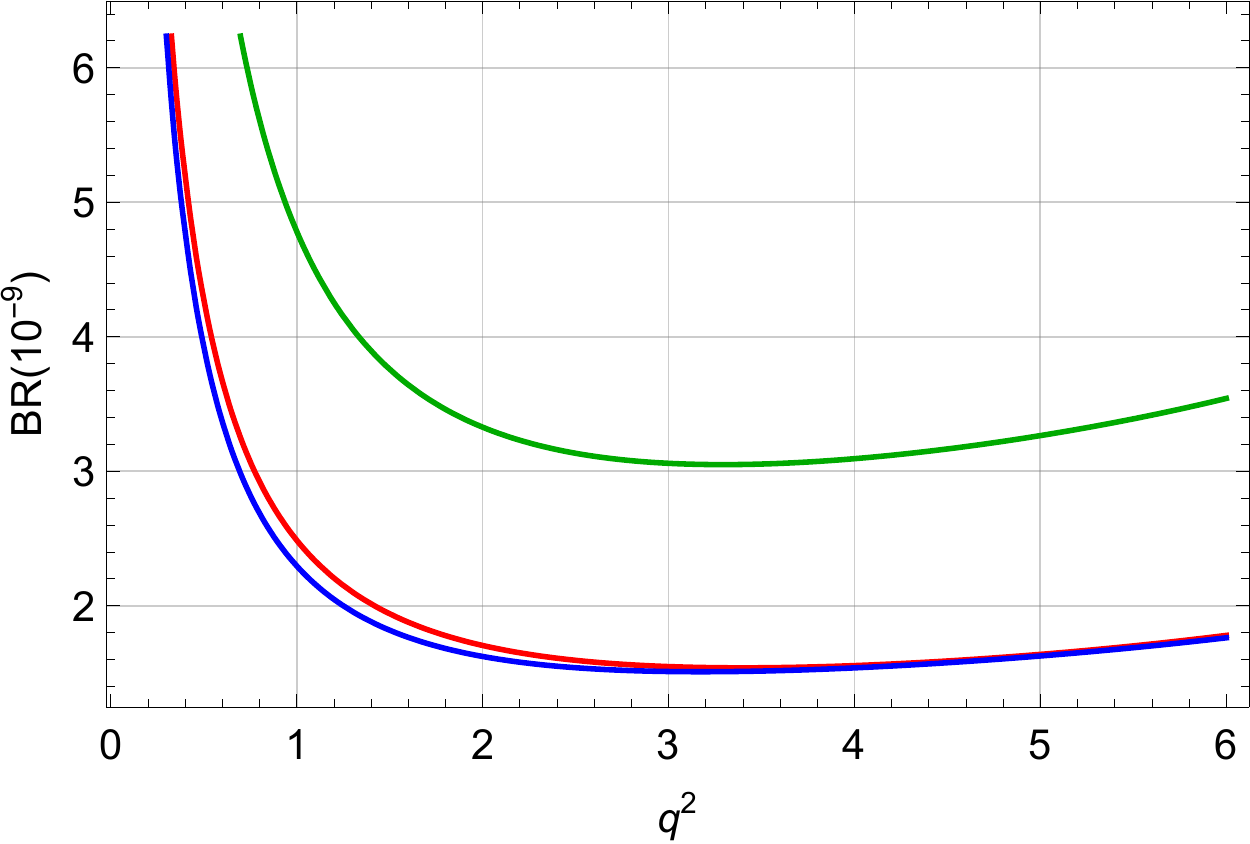}\\
		(g)& (h)\\
		\includegraphics[scale=0.5]{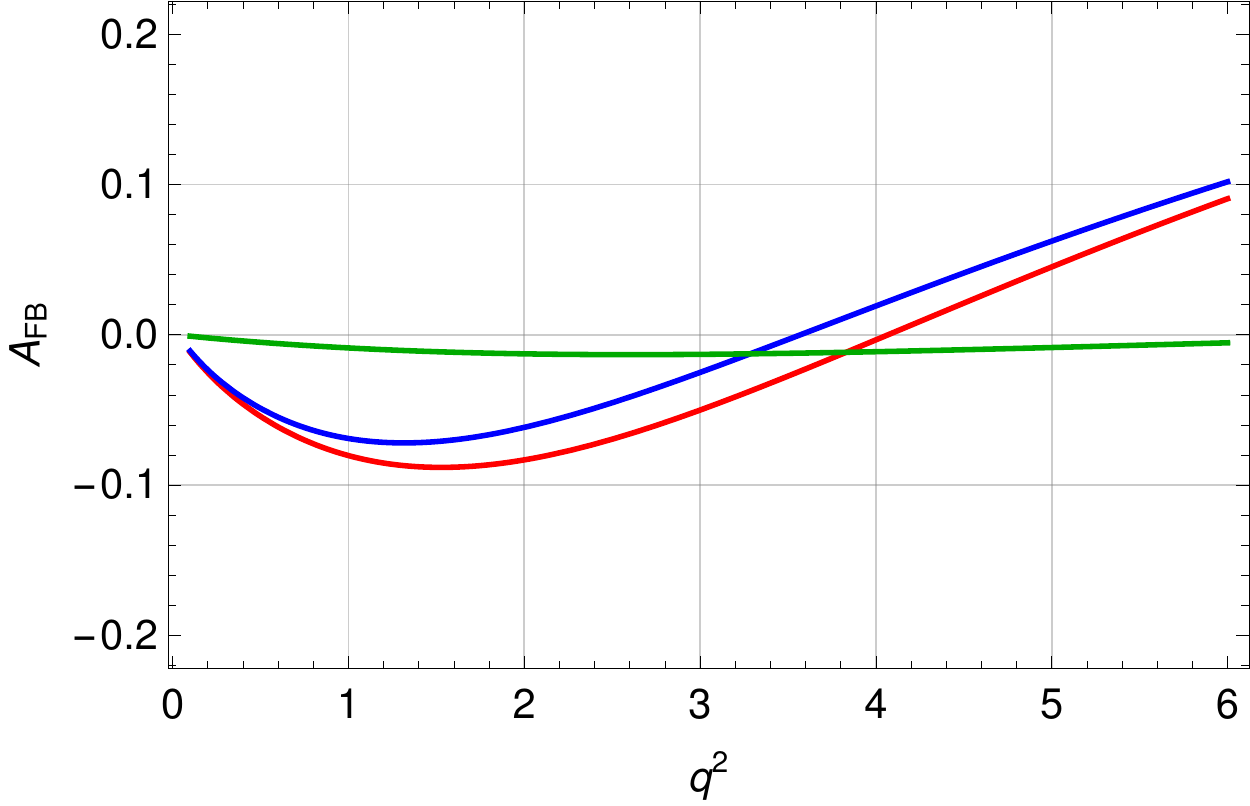}
		~~~~ & ~~~~
		\includegraphics[scale=0.5]{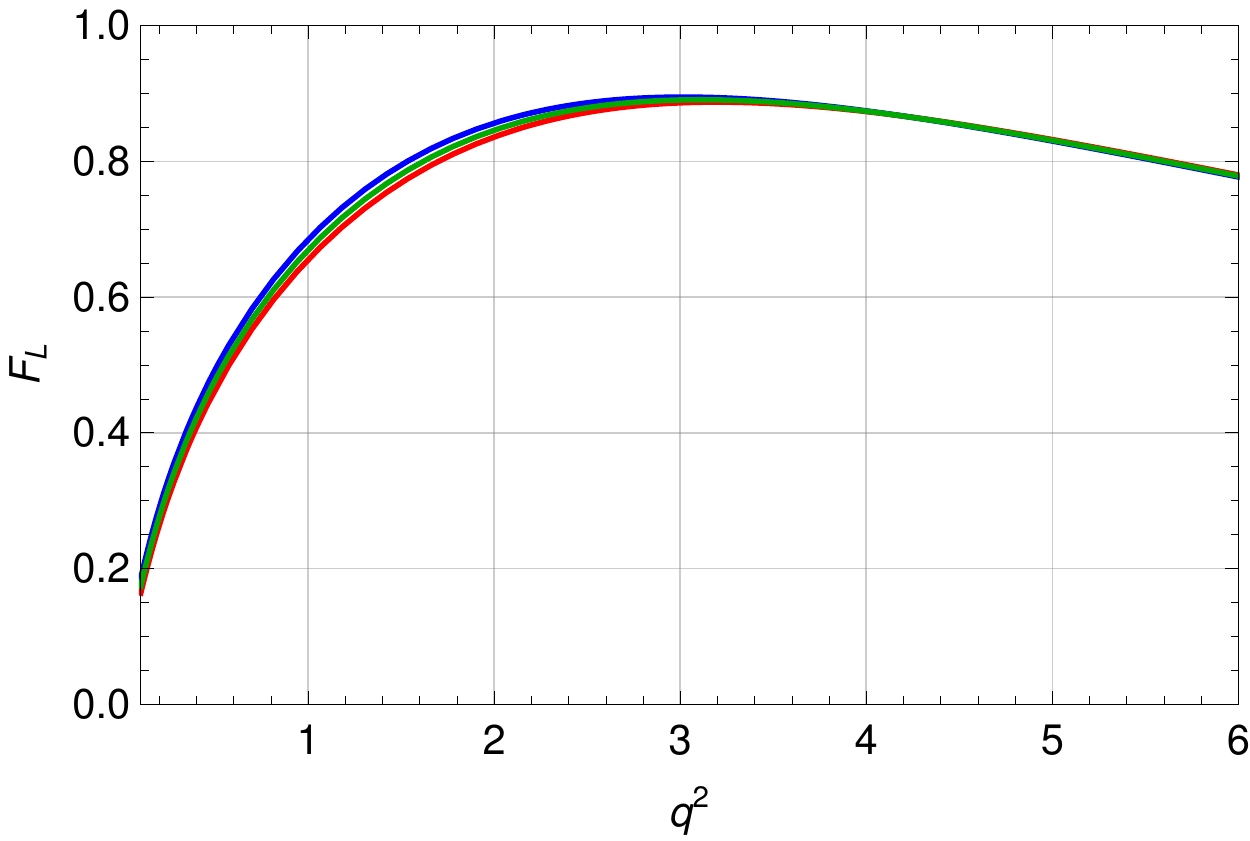}\\
		(i) & (j)
	\end{tabular}
	\caption{\label{rhoLHCbplots} Values of observables to be measured at LHCb as functions of $q^2$. Red curve shows the values for $B\to \rho \mu^+\mu^-$, blue curve shows the values for $\bar{B}\to \rho \mu^+\mu^-$, and Green curve shows the values of observables defined for the untagged events. The plots are obtained using BZ form factors. The mean values include the contribution of non-factorizable corrections.}
\end{figure*}

We now study the impact of various corrections stemming from the four quark operators. The factorizable corrections are already included in the definition of $C_9^{\text{eff}}$ and $C_7^{\text{eff}}$ to NNLO. The non-factorizable ones i.e., weak annihilation, spectator scattering, and soft gluon emission are systematically included for predictions in the bin $[1-6]$GeV$^2$\footnote{Since the parameterization used to define $\Delta C_{9,c}^{\text{soft}}$ are not valid below $1 \text{GeV}^2$, contribution of soft gluon emission has not been included in the lower bin, $[0.1-1]\text{GeV}^2$.}. As mentioned before, the contribution of soft gluon emission from the up quark loop is not available at present. A very rough estimate leads us to include $10\%$ uncertainty in $C_9^{\text{eff}}$ due to this particular correction. The crucial issue here is not just the rough magnitude but also the sign and thus without a proper LCSR based calculation, this is the best one can do. In Tables \ref{KstarCorrec}, \ref{rhoCorrec}, \ref{rhoCorrecLHCb}, we present the value of the observables with these corrections included. In these tables, the first error is due to the form factors while the second shows the spread due to soft gluon emission from the up quark loops. These are presented for the BSZ2 set of form factors. It is found that the inclusion of these corrections has significant impact on observables like $P_5^{\prime}$, branching ratio, and $A_{FB}$. This confirms the broad pattern observed in $B\to K^{\ast}\mu\mu$. Although the results are presented for $[0.1-1]$GeV$^2$ and $[1-6]$GeV$^2$, a comparison of results with and without these corrections is more meaningful and reliable for $[1-6]$ GeV$^2$ bin as for $q^2<1$GeV$^2$, the soft gluon contribution tend to be very large. The observables are also plotted as function of $q^2$ as shown in Fig. \ref{kstarplots},\ref{rhoBelleplots},\ref{rhoLHCbplots}. The values of the zeroes are given in Table \ref{KstarZero},\ref{rhoZeroes}. Since the error due to soft gluon emission from up quark is very small, we only show the error due to form factors in the value of zeroes \footnote{Ignoring the effect of soft gluon emission and time evolution, our results for branching ratio of $B(\bar{B})\to \rho\mu^+\mu^-$ in the bin $[1-6]$GeV$^2$ and zero of $A_{FB}$ are consistent with \cite{Beneke:2004dp}}.\newline
\begin{table*}
	\begin{tabular}{|l||c|c|c|c|}
		\hline
		& \multicolumn{2}{c}{$\bar{B_s}\to K^{\ast}\mu^+\mu^-$} &  \multicolumn{2}{c|}{$B_s\to \bar{K}^{\ast}\mu^+\mu^-$}\\
		\hline
		Observable& [0.1-1] GeV$^2$& [1-6] GeV$^2$&[0.1-1] GeV$^2$& [1-6] GeV$^2$ \\
		\hline
		$P_1$ &$0.012\pm0.129 \pm 0.001$ & $-0.081\pm 0.111\pm 0.005$ & $0.011\pm 0.135\pm 0.001$ & $-0.075\pm 0.108\pm 0.005$\\
		\hline
		
		$P_2$ & $0.118\pm0.013 \pm 0.001$ & $0.112\pm 0.072 \pm 0.036$& $0.112\pm 0.013\pm 0.001$ & $0.142\pm 0.071 \pm 0.034$\\ \hline
		
		$P_3$ & $0.001\pm 0.002 \pm 0.0$ & $0.004\pm0.010\pm 0.002$& $0.001\pm 0.007\pm 0.0$ & $0.003\pm0.010\pm 0.002$ \\  \hline
		
		$P_4^{\prime}$ & $-0.593\pm 0.057 \pm 0.009$ & $0.464 \pm 0.164 \pm 0.014 $ & $-0.650 \pm0.060 \pm 0.008$ & $0.379\pm 0.171 \pm 0.016$\\ \hline
		
		$P_5^{\prime}$ & $0.547\pm0.051 \pm 0.016$& $-0.286\pm 0.125 \pm 0.046$ & $0.543\pm0.053 \pm 0.016$ & $-0.273\pm 0.132 \pm 0.047$\\ \hline 
		
		$P_6^{\prime}$ & $-0.104\pm0.006 \pm 0.016$ & $-0.095\pm 0.011 \pm 0.002$& $-0.069\pm 0.005 \pm 0.001$ & $-0.078\pm 0.004\pm 0.002$\\ \hline
		
		$P_8^{\prime}$ & $0.015\pm0.003\pm 0.016$ & $0.040\pm0.004\pm0.017$ & $0.044\pm0.003\pm 0.016$ & $0.034\pm0.002\pm0.019$\\ \hline
		
		$R_{K^{\ast}}^{B_s}$ &$0.940\pm 0.009 \pm 0.001$& $0.998\pm0.004\pm 0.0$ & $0.942\pm 0.008\pm 0.001$ & $0.998\pm0.004\pm 0.0$\\	\hline
		
			$BR\times 10^{9}$ & $3.812\pm0.450 \pm 0.086$ & $7.803\pm 1.758 \pm 0.357$ & $4.411\pm0.560\pm 0.101$ & $8.391 \pm 1.856\pm 0.375$\\ \hline
		
		$A_{FB}$ & $-0.060 \pm 0.008 \pm 0.001$ & $-0.029 \pm 0.020\pm 0.009$ & $-0.056\pm 0.008\pm 0.001$ & $-0.036\pm 0.020 \pm 0.009 $\\\hline
		
		$F_L$ & $0.453\pm 0.067 \pm 0.014$ & $0.853\pm0.038\pm 0.007$ & $0.464\pm0.064\pm 0.014$ & $0.851\pm0.038\pm 0.007$ \\ \hline
	\end{tabular}
	\caption{\label{KstarCorrec} Observables for $\bar{B_s}\to K^{\ast}\mu^+\mu^-$ and $B_s\to \bar{K^{\ast}}\mu^+\mu^-$} using BSZ2 form factors. The first uncertainty is due to form factors and second is due to soft-gluon corrections with up quark in the loop.
\end{table*}

\begin{table*}
	\begin{tabular}{|c||c|c|}
		\hline
		Observable&$\bar{B}_s\to K^{\ast}\mu^+\mu^-$& $B_s\to \bar{K}^{\ast}\mu^+\mu^-$ \\ \hline
		$P_2$ & $4.137\pm0.421$& $4.307\pm0.441$ \\
		$P_4^{\prime}$& $1.867\pm0.300$ & $2.067\pm0.327$ \\
		$P_5^{\prime}$ & $2.223\pm0.319$ & $2.267\pm0.343$\\
		$A_{FB}$ & $4.081\pm0.453$ & $4.250\pm0.476$ \\
		\hline
	\end{tabular}
\caption{\label{KstarZero} Values of zeroes of angular observables for the process $\bar{B_s}\to K^{\ast}\mu^+\mu^-$ and $B_s\to \bar{K^{\ast}}\mu^+\mu^-$. The uncertainty is due to form factors. Mean values include the contribution of non-factorizable corrections.}
\end{table*}

\begin{table*}
	\begin{tabular}{|l||c|c|c|c|}
		\hline
		& \multicolumn{2}{c}{$B\to\rho\mu^+\mu^-$} &  \multicolumn{2}{c|}{$\bar{B}\to\rho\mu^+\mu^-$}\\
		\hline
		& [0.1-1] GeV$^2$& [1-6] GeV$^2$&[0.1-1] GeV$^2$& [1-6] GeV$^2$\\
		\hline
		$\left<P_1\right>$(Belle)& $0.011\pm0.181 \pm 0.001$ & $-0.059\pm0.110\pm 0.003$ & $0.112\pm0.179\pm 0.001$ & $-0.063\pm 0.113\pm 0.003$  \\
		(LHCb)& $0.050\pm0.181$ & $-0.044\pm 0.110$ & $-0.034\pm 0.179$ & $-0.061\pm 0.111$\\ \hline
		
		$\left<P_2\right>$(Belle) & $0.083\pm0.010\pm 0.001$ & $0.073\pm0.053\pm 0.023$ & $0.008 \pm 0.009\pm 0.0$ & $0.0105\pm 0.050\pm 0.024$ \\
		(LHCb)& $0.083\pm0.010$ & $0.074\pm0.053$ & $0.078 \pm 0.009$ & $0.002\pm 0.050$\\ \hline
		
		$\left<P_3\right>$(Belle)& $0\pm0.005\pm 0.0$ & $0.001\pm0.005\pm 0.002$ & $0\pm0.001\pm 0.0$& $0.001\pm 0.005\pm 0.002$\\
		(LHCb)& $-0.228\pm0.044$ & $-0.229\pm 0.028$ & $0.261\pm 0.050$ & $0.240\pm 0.030$ \\ \hline
		
		$\left<P_4^{\prime}\right>$(Belle)& $-0.618\pm0.076\pm 0.047$ & $0.467\pm0.161\pm 0.029$ & $-0.586\pm 0.076\pm 0.046$ & $0.529\pm 0.155\pm0.017$ \\
		(LHCb) & $-0.591\pm 0.077$ & $0.470\pm 0.161$ & $-0.616\pm 0.075$ & $0.526\pm 0.155$\\ \hline 
		
		$\left<P_5^{\prime}\right>$(Belle) & $0.332\pm0.043\pm0.027$ & $-0.211\pm 0.084\pm0.085$ & $0.300\pm0.040\pm0.030$ & $-0.263\pm0.075\pm0.098$  \\
		(LHCb)& $0.368\pm 0.043$ & $-0.178\pm 0.084$ & $0.331\pm 0.039$  & $-0.228\pm 0.075$\\ \hline
		
		$\left<P_6^{\prime}\right>$(Belle) & $-0.050\pm 0.004 \pm 0.001$ & $-0.064 \pm 0.004 \pm 0.002$ & $-0.088\pm0.006\pm0.001$ & $-0.076\pm0.009\pm0.002$	\\
		(LHCb)& $-0.030\pm 0.003$  & $-0.042\pm 0.003$ & $-0.109\pm 0.008$ & $-0.099\pm 0.010$\\ \hline
		
	$\left<P_8^{\prime}\right>$(Belle) & $0.013\pm 0.002\pm0.016$ & $0.012\pm 0.001\pm0.018$& $0.014\pm 0.005\pm0.002$& $0.020\pm0.003\pm0.017$	 \\
	(LHCb)& $-0.133\pm 0.021$  & $0.113\pm 0.013$ & $-0.145\pm 0.024$ & $0.124\pm 0.014$\\ \hline
		
		$\left<R_{\rho}\right>$(Belle)& $0.938\pm0.203\pm 0.001$& $0.997\pm0.278\pm0.0$& $0.939\pm0.167\pm0.002$ & $0.998\pm 0.362\pm0.0$ \\ 
		(LHCb)& $0.955\pm0.194$ & $1.036\pm0.289$ & $0.954\pm0.192$ & $1.033\pm0.289$\\ \hline
		
		$\left<BR\right>\times 10^{9}$(Belle)& $4.078 \pm 0.585\pm0.080$ & $7.908\pm 1.549 \pm0.366$ & $ 3.653\pm 0.529\pm 0.077$ & $7.626\pm 1.504\pm 0.365$ \\
		(LHCb)& $2.165\pm 0.302$ & $4.064\pm 0.778$ & $1.977\pm 0.273$ & $3.943\pm 0.756$\\ \hline
		
		$\left<A_{FB}\right>$(Belle) & $-0.045 \pm 0.005\pm0.001$ & $-0.023\pm0.018\pm0.007$ & $-0.041\pm 0.005\pm0.001$& $-0.003\pm 0.016\pm0.006$ \\
		(LHCb) & $-0.046\pm 0.005$ & $-0.024\pm 0.018$ & $-0.041\pm 0.001$ & $-0.011\pm 0.002$\\ \hline
		
		$\left<F_L\right>$(Belle)& $0.414\pm0.067\pm 0.014$ & $0.822\pm 0.039\pm 0.007$ & $0.431\pm 0.069 \pm 0.014$ & $0.832\pm 0.038 \pm 0.006$ \\ 
		(LHCb)& $0.409\pm 0.067$ & $0.822\pm 0.039$ & $0.437\pm 0.068$ & $0.832\pm 0.037$ \\ \hline
	\end{tabular}
	\caption{\label{rhoCorrec}Binned values of observables for the process $B\to\rho\mu^+\mu^-$ and $\bar{B}\to\rho\mu^+\mu^-$ using tagged events to be measured at Belle and LHCb. The mean values include non-factorizable corrections. The first uncertainty is due to form factors and second uncertainty is due to soft gluon emission with up quark in the loop.}
\end{table*}

\begin{table*}
	\begin{tabular}{|l||c|c|}
		\hline
		& \multicolumn{2}{c|}{$B\to\rho\mu^+\mu^-$(LHCb)}\\ \hline
		observables & [0.1-1] GeV$^2$& [1-6] GeV$^2$\\ \hline
		$\left<P_1\right>$ & $0.011\pm0.180\pm 0.001 $ & $-0.061\pm0.111 \pm 0.003$\\
		$\left<P_2\right>$& $0.008\pm 0.001\pm 0.0$& $0.033\pm 0.003\pm 0.001$\\
		$\left<P_3\right>$ 	& $0\pm0.002\pm 0.0$ & $0\pm0.001\pm 0.0$\\
		$\left<P_4^{\prime}\right>$& $-0.603\pm0.076\pm 0.008$ & $0.497\pm0.158\pm0.015$\\
		$\left<P_5^{\prime}\right>$& $0.032\pm0.004\pm0.001$ & $0.022\pm0.006\pm0.002$\\
		$\left<P_6^{\prime}\right>$ & $-0.068\pm0.005\pm 0.001$ & $-0.070\pm0.006\pm 0.002$\\
		$\left<P_8^{\prime}\right>$ & $0\pm0.001\pm 0.005$ & $-0.003\pm0.002\pm 0.006$\\
		$\left<R_{\rho}\right>$& $0.954\pm0.193\pm0.002$ & $1.035\pm 0.289\pm0.0$ \\
		$\left<BR\right>\times 10^{9}$& $4.142\pm 0.575\pm0.138$ & $8.007\pm 1.533\pm0.731$\\
		$\left<A_{FB}\right>$ & $-0.046\pm0.005\pm0.0$ & $-0.024\pm0.018\pm0.001$\\
		$\left<F_L\right>$	& $0.422\pm0.068\pm0.032$ & $0.827\pm0.0388\pm0.016$\\
		\hline
	\end{tabular}
	\caption{\label{rhoCorrecLHCb}Binned values for observables for the process $B\to\rho\mu^+\mu^-$ using untagged events to be measured at LHCb. The results include non-factorizable corrections. The first uncertainty is due to form factors and second uncertainty is due to soft gluon emission with up quark in the loop.}	
\end{table*}

\begin{table*}
	\begin{tabular}{|c||c|c|c|c|c|}
		\hline
		Observable& \multicolumn{2}{|c|}{Belle}& \multicolumn{3}{|c|}{LHCb} \\ \hline
		&$B\to \rho\mu^+\mu^-$ & $\bar{B}\to \rho\mu^+\mu^-$ & $B\to \rho\mu^+\mu^-$ & $\bar{B}\to \rho\mu^+\mu^-$ & untagged events\\ \hline
		
		$P_2$ & $4.101\pm 0.441$ & $3.593\pm0.304$ & $4.111\pm0.443$ & $3.604\pm 0.399$ & $-$ \\
		$P_4^{\prime}$ & $1.869\pm 0.304$ & $1.727\pm0.282$ & $1.851\pm 0.307$ & $1.746\pm0.279$ & $1.799\pm0.293$\\
		$P_5^{\prime}$ & $2.107\pm0.344$ & $1.842\pm0.290$ & $2.269\pm0.353$ & $1.860\pm0.287$ & $-$\\
		$P_8^{\prime}$ & - & - & $1.827\pm0.023$ & $1.706\pm 0.036$ & -\\
		$A_{FB}$ & $4.060\pm0.462$ & $3.560\pm0.419$ &$4.069\pm0.462$ & $3.571\pm0.420$ & -\\ \hline
	\end{tabular}
\caption{\label{rhoZeroes} Values of zeroes of angular observables for the process $B\to\rho\mu^+\mu^-$ and $\bar{B}\to\rho\mu^+\mu^-$ The uncertainty is due to form factors. Mean values include the contribution of non-factorizable corrections.}
\end{table*}

\section{ Summary and Conclusions}\label{summary}
Exclusive semileptonic decays mediated by $b\to s$ transitions have shown several deviations from SM expectations \cite{Descotes-Genon:2013wba}. This has attracted a lot of theoretical attention, attempting to explain these deviations. At present, it is not clear if the deviations are due to the physics beyond SM or just hadronic artefacts \cite{charm2,charm4,Ciuchini:2015qxb}. An obvious solution is to study the analogous $b\to d $ transitions. Due to the complex phase involved, $b\to d$ transitions have a rich phenomenology and the CKM parameters $\rho$ and $\eta$ can be extracted from a dedicated study of angular observables \cite{Beneke:2004dp,Khodjamirianpheno,next}. We have provided the detailed predictions of angular observables for $\bar{B}^0\to \bar{\rho}^0\mu^+\mu-$ and $\bar{B}_s\to K^{\ast}\mu^+\mu^-$ modes in the SM. It is found that the naive guess that branching ratio of $\bar{B}^0\to\rho^0\mu^+\mu^-$ should be approximately half of the branching ratio of $\bar{B}_s\to K^{\ast}\mu^+\mu^-$ does not always work because of additional effects due to $B^0-\bar{B}^0$ mixing. It may be worthwhile to mention that the values of observables for $B^{\pm}\to \rho^{\pm}\mu^+\mu^-$ are also expected to be same as that for $B_s(\bar{B_s})\to \bar{K}^{\ast}(K^{\ast})\mu^+\mu^-$ modulo strange quark mass and $SU(3)$ corrections. For the $\bar{B}_s\to K^{\ast}\mu+\mu^-$ mode, we have explicitly compared predictions of angular observables using different form factors. The results, provided in Table \ref{KstarwoCorrec}, clearly show a dependence on the form factors, even for the form factor independent observables, though the deviations in these observables are only marginal. Table \ref{KstarCorrec} contain the predicted values for BSZ2 form factors after including various non-factorizable corrections. These turn out to be important and should be included while comparing with data and to decipher any possible new physics. \newline

A potentially important missing piece is the inclusion of finite width effects, especially relevant for $B\to \rho\ell\ell$ modes. Since, $\rho^0\to\pi\pi$ width is large, it must be taken into account. In \cite{widtheffects}, an attempt is made to include these effects as a part of form factors. However, these effects are computed only for vector and axial vector form factors while no calculation exists for tensor form factors. These effects could be large and must be evaluated. 

To the best of our knowledge, this is the first dedicated study in this direction. The $b\to d$ mediated modes bring along several interesting features due to intrinsic CP violating phase within SM with a non-negligible contribution. We have provided $q^2$ dependence as well as binned values of angular observables, which can be directly compared once the data is available. LHCb's result for $B_s\to \bar{K}^{\ast}\mu^+\mu^-$ \cite{LHCbbs} announced recently is consistent with our result of branching ratio. Precise measurements of various angular observables will lead to complimentary information to $b\to s$ mediated decays. \\
\appendix

\section{Transversity Amplitudes}\label{appendixAmplitudes}

\begin{subequations}
	The non-factorizable corrections discussed in Section can be added to transversity amplitudes or the Wilson coefficient $C_9^{\text{eff}}$. Following \cite{charmloop,Beneke:2001at}, we add the corrections in the following way:
	\begin{align}
	A_{\perp L,R}&(q^2)= \sqrt{2 \lambda} ~N\big[2\frac{m_b}{q^2}(C_7^{\text{eff}}T_1(q^2)+\Delta T_{\perp})\nn\\&  +(C_9^{\text{eff}}\mp C_{10}+ \Delta C_9^{1}(q^2))\frac{V(q^2)}{M_B+M_V} \big]
	\end{align}
	\begin{align}
	A_{\parallel L,R}&(q^2)= -\sqrt{2}N(M_B^2-M_V^2)\big[2\frac{m_b}{q^2}(C_7^{\text{eff}}T_2(q^2)\nn\\&+2 \frac{E(q^2)}{M_B}\Delta T_{\perp})+ (C_9^{\text{eff}}\mp C_{10}+\Delta C_9^2(q^2))\nn\\&\frac{A_1(q^2)}{M_B-M_V}\big]
	\end{align}
	\begin{align}
	A_{0 L,R}&(q^2)=-\frac{N}{2M_V \sqrt{q^2}}\big[2 m_b\big((M_B^2+3M_V^2-q^2)\nn\\ &(C_7^{\text{eff}}T_2(q^2)\big)-\frac{\lambda}{M_B^2-M_v^2}(C_7^{\text{eff}}T_3(q^2)+\Delta T_{\parallel}))\nn\\&+(C_9^{\text{eff}}\mp C_{10}+\Delta C_9^3)\nn \\&\big((M_B^2+M_V^2-q^2)(M_B+M_V)A_1(q^2)\nn\\&-\frac{\lambda}{M_B+M_V}A_2(q^2)\big)\big]
	\end{align}
	\begin{align} 
	A_t&(q^2)=\frac{N}{\sqrt{s}}\sqrt{\lambda}2 C_{10}A_0(q^2)
	\end{align}
\end{subequations}
where, 
\begin{align}
\Delta T_{\perp}&=\frac{\pi^2}{N_c} \frac{f_{P}f_{V,\perp}}{M_B}\frac{\alpha_sC_F}{4\pi}\int\frac{d\omega}{\omega}\Phi_{P,+}(\omega)\nn\\ &\int_0^1du~ \Phi_{V,\perp}(u)(T_{\perp}^{c,\text{spec}}+\frac{\xi_u}{\xi_t}(T_{\perp}^{u,\text{spec}}))
\end{align}
\begin{align}
\Delta T_{\parallel}&=\frac{\pi^2}{N_c} \frac{f_{P}f_{V,\parallel}}{M_B}\frac{M_V}{E}\sum_{\pm}\int\frac{d\omega}{\omega}\Phi_{P}(\omega)\nn\\ &\int_0^1du~ \Phi_{V,\parallel}(u)\big[T_{\parallel}^{c,WA}+\frac{\xi_u}{\xi_t}T_{\parallel}^{u,WA}+\nn \\&
\frac{\alpha_sC_F}{4\pi}(T_{\parallel}^{c,\text{spec}}+\frac{\xi_u}{\xi_t}T_{\parallel}^{,\text{spec}})\big]\\
\Delta C_9^i&=\Delta C_{9,c}^{i,soft}+\Delta C_{9,u}^{i,soft}
\end{align}
where $P\equiv \bar{B},\bar{B}_s$ and $V\equiv \rho,K^{\ast}$. The values of input parameters used to calculate the corrections are given in Tables \ref{input} and \ref{input2}. 

\section{Observables}\label{appendixA}
In this section, we explicitly write the definitions of observables considered while giving predictions for $B\to \rho\mu^+\mu^-$ process.
\subsection{Tagged}\label{tagged}
The observables for $B^0\to \rho^0\mu^+\mu^-$ corresponding to tagged events which can be measured at LHCb and Belle are defined as, 
\begin{eqnarray}\label{belleob}
	\left<\frac{d\Gamma}{dq^2}\right>^{\text{Tagged}}=&\frac{1}{4}(3J_1^c+6J_1^s-J_2^c-2J_2^s)\\
	\left<A_{FB}(q^2)\right>^{\text{Tagged}}= &\frac{-3J_6^s}{3J_1^c+6J_1^s-J_2^c-2J_2^s}\\
	\left<F_L(q^2)\right>^{\text{Tagged}}=&\frac{3J_1^c-J_2^c}{3J_1^c+6J_1^s-J_2^c-2J_2^s} 
	\\
	\left<R_{\rho}\right>^{\text{Tagged}}=&\frac{\int_{q_1^2}^{q_2^2}~dq^2~\left<d\Gamma/dq^2\right>^{\text{Tagged}}}{\int_{q1^2}^{q_2^2}~dq^2~\left<d\Gamma/dq^2\right>^{\text{Tagged}}}
\end{eqnarray}
\begin{eqnarray}
	\left<P_1\right>^{\text{Tagged}}=&\frac{J_3}{2 J_{2s}},& \left<P_2\right>^{\text{Tagged}}=\beta_l\frac{J_6^s}{8J_2^s},\nn \\
	\left<P_3\right>^{\text{Tagged}}=&\frac{J_9}{4J_2^s},&
	\left<P_4^{\prime}\right>^{\text{Tagged}}=\frac{J_4}{\sqrt{-J_2^cJ_2^s}},\nn\\ \left<P_5^{\prime}\right>^{\text{Tagged}}=&\frac{J_5}{2\sqrt{-J_2^cJ_2^s}},&
	\left<P_6^{\prime}\right>^{\text{Tagged}}=\frac{-I_7}{2\sqrt{-I_2^cI_2^s}},\nn\\ 
	\left<P_8^{\prime}\right>^{\text{Tagged}}=&\frac{-I_8}{2\sqrt{-I_2^cI_2^s}}&\\
	 \label{belleob2}
	\end{eqnarray}
The observables for the CP-conjugate decay $\bar{B}^0\to \rho\mu^+\mu^-$ are obtained by replacing $J_i$ by $\tilde{J}_i(\equiv\zeta_i \bar{I}_i)$ in Eqs. (\ref{belleob}-\ref{belleob2}). These definitions are common for observables at LHCb and Belle, but the definitions of angular functions are different for the two cases. For Belle, the functions $J_i$ and $\tilde{J}_i$ used are time integrated functions obtained from Eq. (\ref{modified}) and given as,
\begin{eqnarray}
J_i&=\frac{1}{\Gamma}[I_i +\tilde{I}_i+\frac{1}{1+x^2}\times ( I_i +\tilde{I}_i])\nn\\
\tilde{J}_i&=\frac{1}{\Gamma}[I_i +\tilde{I}_i-\frac{1}{1+x^2}\times (I_i +\tilde{I}_i)]
\end{eqnarray}
while for LHCb, the angular functions are given by, 
\begin{eqnarray}
J_i&=\frac{1}{2\Gamma}[I_i +\tilde{I}_i+\frac{1}{1+x^2}\times( I_i +\tilde{I}_i)-\frac{x}{1+x^2}\times s_i]\nn\\
\tilde{J}_i&=\frac{1}{2\Gamma}[I_i +\tilde{I}_i-\frac{1}{1+x^2}\times (I_i+\tilde{I}_i) +\frac{x}{1+x^2}\times s_i]
\end{eqnarray}
\subsection{Untagged}\label{Untagged}
For untagged events, the observables for $B^0\to \rho^0\mu^+\mu^-$ are defined as, 
\begin{eqnarray}\label{lhcbob}
	\left<\frac{d\Gamma}{dq^2}\right>^{\text{Untagged}}=&\frac{1}{2}	\left<\frac{d\Gamma}{dq^2}+\frac{d\bar{\Gamma}}{dq^2}\right>^{\text{Untagged}}\\
	\left<A_{FB}(q^2)\right>^{\text{Untagged}}= &\frac{-3(J_6^s+\tilde{J}_6^s)}{4\left<d\Gamma/dq^2\right>^{\text{Untagged}}}\\
	\left<F_L(q^2)\right>^{\text{Untagged}}=&\frac{3(J_1^c+\tilde{J}_1^c)-(J_2^c+\tilde{J}_2^c)}{4\left<d\Gamma/dq^2\right>^{\text{Untagged}}} 
\end{eqnarray}
\begin{equation}
	\left<R_{\rho}\right>^{\text{Untagged}}=\frac{\int_{q_1^2}^{q_2^2}~dq^2~\left<d\Gamma/dq^2\right>^{\text{Untagged}}+\left<d\Gamma/dq^2\right>^{\text{Untagged}}}{\int_{q1^2}^{q_2^2}~dq^2~\left<d\Gamma/dq^2\right>^{\text{Untagged}}+\left<d\Gamma/dq^2\right>^{\text{Untagged}}}
	\end{equation}

\begin{subequations}\label{lhcbob}
	\begin{align}
	\left<P_1\right>^{\text{Untagged}}=&\frac{J_3+\tilde{J}_3}{2 (J_{2s}+\tilde{J}_2^s)},\\ \left<P_2\right>^{\text{Untagged}}=&\beta_l\frac{J_6^s+\tilde{J}_6^s}{8(J_2^s+ \tilde{J}_2^s)},\\
	\left<P_3\right>^{\text{Untagged}}=&\frac{J_9+\tilde{J}_9}{4(J_2^s+\tilde{J}_2^s},\\
	\left<P_4^{\prime}\right>^{\text{Untagged}}=&\frac{J_4+\tilde{J}_4}{\sqrt{-(J_2^c+\tilde{J}_2^c)(J_2^s+\tilde{J}_2^s)}},\\ \left<P_5^{\prime}\right>^{\text{Untagged}}=&\frac{J_5+\tilde{J}_5}{2\sqrt{-(J_2^c+\tilde{J}_2^c)(J_2^s+\tilde{J}_2^s)}},\\
	\left<P_6^{\prime}\right>^{\text{Untagged}}=&\frac{-(J_7+\tilde{J}_7)}{2\sqrt{-(J_2^c+\tilde{J}_2^c)(J_2^s+\tilde{J}_2^s)}}\\
	\left<P_8^{\prime}\right>^{\text{Untagged}}=&\frac{-(J_8+\tilde{J}_8)}{2\sqrt{-(J_2^c+\tilde{J}_2^c)(J_2^s+\tilde{J}_2^s)}}.
	\end{align}
\end{subequations}

  In section \ref{tagged}, all the $J_i$ and $\tilde{J}_i$ are time integrated functions. Similarly in section \ref{Untagged}, the combination $J_i+\tilde{J}_i$ are time integrated functions and the $\left<\right>$ symbol is suppressed. 
 
\section*{Acknowledgement}
 Authors thank S. Mohanty for making the computational resources of his TDP projects available for a large part of numerical analysis carried out in this paper. Authors also thank J. Virto, T. Gershon, and T. Blake for their useful comments and suggestions.

\end{document}